\documentstyle[seceq,epsf]{ptptex}

\def\eg{{\it e.g.}\ }

\def\etal{{\it et\ al.\ }}

\newcommand{\lsim}
 {\ \raise.35ex\hbox{$<$}\kern-0.75em\lower.5ex\hbox{$\sim$}\ }
\newcommand{\gsim}
 {\ \raise.35ex\hbox{$>$}\kern-0.75em\lower.5ex\hbox{$\sim$}\ }
\def\journal #1#2#3#4{#1 {\bf #2} (#4) #3}
\def\PR{Phys.\ Rev.}
\def\PRB{Phys.\ Rev.\ B}
\def\PRL{Phys.\ Rev.\ Lett.}

\def\JLTP{J.~Low Temp.~Phys.}

\def\JPCM{J.\ Phys.\ Cond.\ Mat.}
\def\JPCS{J.\ Phys.\ Chem.\ Solids}

\def\JPSJ{J.\ Phys.\ Soc.\ Jpn.}
\def\MPLB{Mod.\ Phys.\ Lett.\ B}

\def\RMP{Rev.\ Mod.\ Phys.}
\def\PTP{Prog.\ Theor.\ Phys.}
\def\ZP{Z.\ Phys.}
\def\ZPB{Z.\ Phys.\ B}
\title{
Variational Monte Carlo studies of attractive Hubbard model I
}
\subtitle{Normal-state properties}

\author{
Hisatoshi {\sc Yokoyama} 
\footnote{E-mail address: yoko@cmpt.phys.tohoku.ac.jp}
}

\inst{
Department of Physics, Tohoku University, Aramaki Aoba, Sendai 980-8578\\
}

\recdate{
}

\abst{
Normal states of the attractive Hubbard model, especially in two 
dimension, are studied in the light of a transition from a Fermi liquid 
to an insulating or gapped state. 
A series of variational Monte Carlo calculations with better statistics 
is carried out to estimate accurately expectation values by several 
many-body wave functions. 
Although a relatively clear crossover is observed even in the plain 
Gutzwiller wave function, the states in both regimes are metallic. 
Meanwhile, a substantial metal-insulator transition takes place at 
$|U|\sim W$ (band width) in an improved wave function in which intersite 
correlation is introduced by taking account of virtual states in 
the second-order perturbation in the infinite-$|U|$ limit. 
The critical value is favorably compared with recent results of the 
dynamical-mean-field approximation. 
In contrast, a conventional Jastrow-type wave functions scarcely 
improve the normal state. 
In addition, the issue of Brinkman-Rice metal-insulator transition is 
reconsidered with much larger systems. 
}

\begin{document}

\maketitle

\section{Introduction} 
To consider the pseudogap phenomena found in cuprate superconductors, 
knowledge as to the attractive Hubbard model (AHM) is useful especially 
in the following points: 
i) AHM undergoes a crossover of the mechanism of superconductivity 
from a BCS type to a Bose-condensation one with increasing the 
correlation strength.\cite{AI} 
ii) Controllable and quantitative calculations are relatively easy. 
iii) Through a canonical transformation AHM on typical bipartite lattices 
connects with the repulsive Hubbard model (RHM)---a plausible model 
for the high-$T_{\rm c}$ cuprates in two dimension (2D). 
To understand the item i) as well as pseudogap, it is indispensable to 
study normal-state properties in low temperature, even though an s-wave 
superconducting phase prevails in the whole ground-state phase diagram 
of AHM in higher lattice dimensions.\cite{Micnus,Scalettar} 
So far, the above crossover and accompanying anomalous or gapped normal-state 
properties, for instance, in static spin susceptibility and single-particle 
spectral function have been studied by such approaches as quantum Monte 
Carlo (QMC) methods,\cite{TR,Singer} $T$-matrix approximations.\cite{TM} 
Quite recently it was shown by dynamical mean-field approximations 
(DMFT)\cite{MetznerDMF,Capone} that the crossover from a Fermi liquid 
to a gapped state at finite temperature is linked to a metal-insulator 
transition, which arises at $|U|\sim W$ ($W$: band width), in low 
temperature.\cite{noteexp} 
\par
 
Thanks to these studies, our understanding on the normal state of AHM 
has been promoted. 
However, many of the above results are not conclusive, because each of 
the above approaches has weak points. 
$T$-matrix approximations are reliably applied only to a low-density regime 
of weak to intermediate correlation strength. 
The validity of DMFT is not confirmed for realistic dimensions, particularly 
in two dimension (2D), and the information on spatial correlation is 
not drawn. 
As for QMC methods, it is not easy to pursue a normal state for 
$T<T_{\rm c}$. 
In addition, the reliability deteriorates with increasing correlation 
strength beyond the band width as well as the system size,\cite{Guber} 
although the minus sign problem is absent. 
Thus, it is necessary to corroborate those findings by complementary 
methods.
\par 

The purpose of this paper is to study the normal-state properties 
of AHM by applying a variational Monte Carlo (VMC) method directly to 
2D systems. 
Merits of the VMC approach lie in the applicability to any correlation 
strength and any electron density, to a variety of many-body trial 
wave functions, and to large enough lattices to check system-size 
dependence. 
Taking advantage of these merit, we especially lay emphasis on crossovers 
or metal-insulator transitions of the states with the above situation 
in mind. 
As another aim we consider the resemblance and difference between AHM 
and RHM, which are interconnected by a canonical transformation. 
Such a check is indispensable before a direct comparison of the results 
of AHM with the experiments of the high-$T_{\rm c}$ cuprates. 
\par

This paper is organized as follows: 
In \S2 the model is defined, and the relationship between the 
attractive and the repulsive Hamiltonians is briefly described. 
Thereby, the previous results of RHM sometimes become applicable 
to AHM. 
In \S3 the well-known Gutzwiller wave function (GWF) is studied 
to show a metal-to-metal crossover between weak- and strong-correlation 
regimes. 
In this connection the absence of Brinkman-Rice transition, which 
arises when one resorts to an additional mean-field-type approximation 
(so-called Gutzwiller approximation) in estimating expectation values, 
is checked again based on the VMC data in 1D, 2D and 3D with better 
statistics in Appendix A. 
Furthermore, in Appendix B we confirm the absence of long-range orders 
within GWF. 
In \S4 to improve the normal state on GWF we introduce a binding 
factor between $\uparrow$- and $\downarrow$-spin electrons by taking 
account of perturbation processes in the strong coupling limit. 
This wave function not only remedies the shortcomings of GWF but 
undergoes a substantial metal-insulator transition. 
As a supplement to \S4 details of the metal-insulator transition in 
this improved wave function are described in Appendix D. 
As an alternative improvement on GWF an distance-dependent intersite 
correlation (we call `Jastrow') factor are studied in Appendix C. 
Section 5 is assigned to summary. 
\par

\section{Model and Method}

\subsection{Attractive Hubbard model} 
In this paper we study a couple of basic wave functions as a normal 
state for the attractive Hubbard model ($U\le 0$): 
\begin{equation}
{\cal H}={\cal H}_t+{\cal H}_U
   =-t\sum_{\langle i,j\rangle\sigma}
   \left(c^\dagger_{i\sigma}c_{j,\sigma}
        +c^\dagger_{j\sigma}c_{i,\sigma}\right)
   +U\sum_jn_{j\uparrow}n_{j\downarrow},
\label{eq:chamil}
\end{equation}
where $n_{j\sigma}=c_{j\sigma}^\dagger c_{j\sigma}$ and other notations 
are standard. 
We use $t$ as the unit of energy.
In this paper only bipartite lattices are taken up, and the sum of 
transfer in eq.~(\ref{eq:chamil}) is restricted to the nearest-neighbor 
pairs. 
We concentrate on the 2D square lattice except for Appendix A, where 
the dependence on lattice dimension becomes an important element. 
Due to the electron-hole symmetry we restrict electron density 
$n$ ($=N_{\rm e}/N$) to $n\le 1$ with $N_{\rm e}$ and $N$ being the numbers 
of electrons and sites, respectively. 
Here, we exclude an external field, thus 
$N_\uparrow=N_\downarrow=N_{\rm e}/2$ 
($N_\sigma$: number of $\sigma$-spin electrons). 
A chemical potential term $-\zeta\sum_{j\sigma}n_{j\sigma}$ may be 
added to adjust electron density, if necessary. 
For later use we summarize a couple of basic facts about this model
in the following. 
\par

For bipartite lattices, AHM is related to RHM by a celebrated canonical 
transformation,\cite{Canon,Shiba} 
\begin{equation}
c_{j\uparrow}=b_{j\uparrow}, \quad 
c_{j\downarrow}=e^{iG\cdot r_j}b_{j\downarrow}^\dagger,
\label{eq:canon}
\end{equation}
where $G$ is the antiferro reciprocal lattice vector, \eg $G=(\pi,\pi)$ 
for the 2D square lattice, and $r_j$ is the position of site $j$.
The transformed Hamiltonian is written as,
\begin{equation}
\tilde{\cal H}=-t\sum_{\langle i,j\rangle\sigma}
   \left(b^\dagger_{i\sigma}b_{j,\sigma}
        +b^\dagger_{j\sigma}b_{i,\sigma}\right)
   -U\sum_j\tilde n_{j\uparrow}\tilde n_{j\downarrow}
   +UN\tilde n_\uparrow
   -h\sum_j\left(\tilde S^z_j+\frac{1}{2}\right),
\label{eq:bhamil}
\end{equation}
where $\tilde n_{j\sigma}=b_{j\sigma}^\dagger b_{j\sigma}$, 
$\tilde S^z_j=(\tilde n_{j\uparrow}-\tilde n_{j\downarrow})/2$ and 
$\tilde n_\sigma=\tilde N_\sigma/N$. 
Henceforth, representation with a tilde denotes the transformation 
by eq.~(\ref{eq:canon}). 
Thus, the sign of the interaction term is reversed, and the chemical 
potential $\zeta$ (electron density $n$) in AHM is related to the 
effective magnetic field as $h=2\zeta$ (magnetization as $m=1-n$) in $z$ 
direction in RHM. 
Moreover, unless the original AHM possesses spin polarization, electron 
density in the transformed RHM is always at half filling due to the 
relation,
$\tilde n_{j\uparrow}+\tilde n_{j\downarrow}
=1+n_{j\uparrow}-n_{j\downarrow}$. 
Meanwhile, the order parameters of antiferro CDW and onsite singlet 
pairing defined as, 
$ 
O_{\rm CDW}=\frac{1}{N}\left|\sum_j(-1)^j 
\langle n_{j\uparrow}+n_{j\downarrow}-1\rangle\right|\ 
$ 
and
$ 
O_{\rm SC}=\frac{1}{N}\sum_j 
\langle c_{j\uparrow}^\dagger c_{j\downarrow}^\dagger\rangle\ 
{\rm or}\ \frac{1}{N}\sum_j 
\langle c_{j\downarrow}c_{j\uparrow}\rangle, 
$
are transformed into the forms of $z$ and $x,y$ components of 
SDW order parameter, 
$
O_{\rm SDW}^z=\frac{1}{N}\left|\sum_j(-1)^j 
\langle n_{j\uparrow}-n_{j\downarrow}\rangle\right|\ 
$ 
and
$
O_{\rm SDW}^\pm=\frac{1}{N}\sum_j 
\langle c_{j\uparrow}^\dagger c_{j\downarrow}\rangle\ 
{\rm or}\ \frac{1}{N}\sum_j 
\langle c_{j\downarrow}^\dagger c_{j\uparrow}\rangle, 
$
respectively.\cite{Nagaoka}
Through this transformation one can deduce the properties of AHM 
from the knowledge of RHM, and vice versa. 
In particular, at half filling ($\zeta=0$) the relations become simple 
owing to the disappearance of the effective field.
For example, it is widely accepted that an antiferromagnetic long-range 
order with a common magnitude of $O_{\rm SDW}^\alpha$ among 
$\alpha=x,y,z$ arises in the half-filled RHM on 2D square lattice for 
arbitrary $U\ (>0)$. 
This reads as AHM at half filling simultaneously possesses 
a CDW and a singlet pairing order of the same magnitude.\cite{Nagaoka} 
Note that it is restricted to exact treatments that the above mapping 
holds unconditionally; when some approximation is applied, the validity 
of mapping has to be verified against individual treatments.
\par

Next, let us consider the strong coupling case of eq.~(\ref{eq:chamil}). 
As the $t$-$J$ model is derived from RHM, an effective Hamiltonian 
for large-$|U|/t$ regime is obtained by picking out the lowest-order 
terms in a perturbation expansion in $t/|U|$ as,\cite{Harris,Emery} 
\begin{equation}
{\cal H}_{\rm eff}
=\frac{2t^2}{|U|}\sum_{\langle ij\rangle}
\left(-p_i^\dagger p_j+\rho_i\rho_j\right), 
\label{eq:ehamil}
\end{equation}
where $p_i=c_{i\uparrow}c_{i\downarrow}$ and 
$\rho_i$ [$=(n_{i\uparrow}+n_{i\downarrow})/2$] is the corresponding number 
operator. 
They operate in the subspace with no singly-occupied site. 
Thus, the system is reduced to hard-core bosons with nearest-neighbor 
repulsive interaction. 
Notice that in the strong-correlation region where eq.(\ref{eq:ehamil}) 
is valid, every energy within `$p$'-band is scaled by $t^2/|U|$, 
which is much smaller than the energy of order $|U|$ needed to pull 
a pair apart: the Hubbard gap. 

\subsection{Variational Monte Carlo method}
We use a conventional variational Monte Carlo (VMC) method.\cite{YS1} 
However, the accuracy of calculations in this paper is notably 
improved by virtue of the rapid progress of computer technology in 
the past decade.
In this work we chiefly consider the 2D square lattice with $N=L\times L$ 
sites, and take up the 3D simple cubic lattice with $N=L\times L\times L$ 
only in Appendix A to see space-dimension dependence of GWF. 
To discuss the thermodynamic limit we carefully check the system-size 
dependence with systems up to $N=2N_{\rm e}=28\times28=784$ and 
$N=N_{\rm e}=24\times24=576$ for 2D and 
$N=N_{\rm e}=10\times10\times10=1000$ for 3D. 
\par

The periodic(P)-antiperiodic(A) boundary condition is imposed on all 
the systems we use for the square lattice. 
Thereby, systems with arbitrary even $L$ for $n=1$ can be treated 
satisfying the closed shell structure, which enables us to check system-size 
dependence. 
On the other hand, for $n=0.5$ available systems are limited to 
$L=8,12,20,28$ etc.\ under this condition, but we may regard the system 
of $L=28$ as sufficiently large as will be mentioned later. 
For incommensurate filling it is practically impossible to directly 
check system-size dependence, because available densities are different
size by size. 
For the 3D simple cubic lattice in $n=1$ the P-P-A and the P-A-A boundary 
conditions are used for lattices with $L=4I$ and $L=4I+2$ ($I$: integer), 
respectively, to satisfy the closed shell condition. 
To reduce the statistical error a great number of samples are used: 
$2.5\times 10^5$ to $10^7$ according to $N_{\rm e}$. 
To keep the mutual independence in electron configurations among 
samples, three kinds of trials (single electron hopping, exchange 
and pair hopping) are adopted; besides, the sampling intervals are 
adjusted up to 100 Monte-Carlo steps so that each acceptance ratio 
may reach 50\%, except for some limiting cases. 
\par 

\section{Gutzwiller wave function} 
In this section as a continuation of the previous researches for 
RHM\cite{YS1} we study the properties of the Gutzwiller wave function 
for AHM. 
In \S3.1 we briefly summarize fundamentals of GWF for later use. 
In \S3.2 and \S3.3 we argue the VMC results for energy and correlation 
functions, respectively.
\par

\subsection{General aspects}
As a trial wave function for normal phases of AHM, the celebrated 
GWF,\cite{GWF} 
\begin{equation}
\Psi_{\rm G}={\cal P}_{\rm G}(g)\Phi_{\rm F}
=\prod_j
\left[1-(1-g)n_{j\uparrow}n_{j\downarrow}\right]\Phi_{\rm F} 
=\exp\left(\alpha\sum_jn_{j\uparrow}n_{j\downarrow}\right)\Phi_{\rm F}, 
\label{eq:GWF}
\end{equation}
is again a useful starting point due to its simplicity. 
In eq.~(\ref{eq:GWF}), $g$ or $\alpha (=\ln g)$ is the sole variational 
parameter, which controls the ratio of doubly occupied sites $d$, 
and $\Phi_{\rm F}$ is the Fermi sea. 
For AHM the range of $g$ will be $1\le g\le\infty$ to enhance $d$, 
whereas for RHM $0\le g\le 1$ to reduce $d$. 
The two limiting cases, $g=0$ and $g=\infty$, indicate insulating states 
without doubly-occupied sites and without singly-occupied sites, 
respectively. 
The former state is used for the $t$-$J$ model,\cite{t-J} 
and becomes the exact ground state particularly for the 1D supersymmetric 
model with $1/r^2$ coupling.\cite{KY}
For $g=1$ GWF is reduced to the noninteracting case, $\Phi_{\rm F}$. 
\par

So far, GWF has been extensively studied for RHM to treat itinerant 
magnetism, Mott transition, $^3$He, superconductivity etc. 
Because of the difficulty in calculating the expectation values by GWF, 
they had been estimated by a mean-field-type approximation [Gutzwiller 
approximation (GA)] for a long time since the first introduction by 
Gutzwiller himself.\cite{Gutz,Voll}
Such an additional approximation, however, breaks the variation principle 
and sometimes leads to qualitatively wrong results. 
A typical erroneous case is the Brinkman-Rice metal-insulator 
transition,\cite{BRT} which never takes place actually in finite dimensions. 
We will reconsider this point in detail in Appendix A. 
Afterward, accurate calculations have been carried out by a VMC 
method\cite{YS1} and diagrammatic calculations for $D=1$ and 
$\infty$.\cite{MV,GV} 
As for AHM, the behavior of GWF has not yet been addressed sufficiently, 
although GA has been already applied.\cite{GAA} 
\par

To take advantage of the fruits of RHM, let us look into the relationship 
between GWF's in the two models through the canonical transformation 
eq.~(\ref{eq:canon}), first. 
Since the relations of operators in $k$ space are 
$c_{k\uparrow}=b_{k\uparrow}$ and 
$c_{k\downarrow}=b_{-k+G\downarrow}^\dagger$, 
the vacuum state for $b_\downarrow$ operators is defined as 
$
|\tilde 0\rangle_\downarrow
      =\prod_k c_{k\downarrow}^\dagger|0\rangle_\downarrow
$ 
and inversely 
$ 
|0\rangle=\prod_k b_{k\downarrow}^\dagger|\tilde 0\rangle_\downarrow.\ 
$
Then, the transformed Fermi sea is written as,
\begin{equation}
\tilde\Phi_{\rm F}=\prod_{k<k_{\rm F}}b_{k\uparrow}^\dagger
\prod_{k<k_{\rm F}}b_{k+G\downarrow}\prod_{k'}b_{k'\downarrow}^\dagger
|\tilde 0\rangle.
\label{eq:bFermisea}
\end{equation}
Except for half filling, $\tilde\Phi_{\rm F}$ has a finite magnetization 
according to the effective field $h$ in eq.~(\ref{eq:bhamil}). 
In particular, if the nesting condition is completely satisfied, \eg
for the square lattice and the simple cubic lattice at half filling, 
$\tilde\Phi_{\rm F}$ becomes equivalent to $\Phi_{\rm F}$. 
On the other hand, the Gutzwiller factor is transformed into
$
\tilde{\cal P}_{\rm G}=\exp\left(\alpha N_\uparrow\right) 
\exp\left(-\alpha\sum_j\tilde n_{j\uparrow}\tilde n_{j\downarrow}\right), 
$
due to 
$(\sum_jn_{j\uparrow})\tilde\Phi_{\rm F}=N_\uparrow\tilde\Phi_{\rm F}$. 
Consequently, by neglecting the constant factor, the transformed 
GWF is written as 
\begin{equation}
\tilde\Psi_{\rm G}=
\prod_j\left[1-\left(1-\frac{1}{g}\right)
\tilde n_{j\uparrow}\tilde n_{j\downarrow}\right]\tilde\Phi_{\rm F}.
\label{eq:tildeGWF}
\end{equation}
By comparing $\tilde\Psi_{\rm G}$ with the original GWF eq.~(\ref{eq:GWF}), 
the Gutzwiller parameter $g$ is replaced by $1/g(\equiv\gamma)$, in 
addition to a reversal of the Fermi sea eq.~(\ref{eq:bFermisea}). 
In particular, if the system has the electron-hole symmetry, the relation, 
\begin{equation}
E_t(g)=E_t(\gamma),\quad 
\frac{E_U(g)}{U}=\frac{1}{2}-\frac{E_U(\gamma)}{U}, 
\label{eq:symGWF}
\end{equation} 
is derived from the equivalence between $\Psi_{\rm G}(g)$ and 
$\tilde\Psi_{\rm G}(\gamma)$ and eq.~(\ref{eq:bhamil}).
Here, $E_t(g)$ and $E_U(g)$ are the variational expectation values of 
${\cal H}_t$ and ${\cal H}_U$ with respect to $\Psi_{\rm G}(g)$, 
respectively. 
Thus, the energetics for RHM is simply mapped to the one for AHM 
in this special case. 
In Figs.~\ref{fig:EsymGWF}(a) and (b), $E_t$ and $E_U$ by GWF for the 
square lattice are shown as a function of $\log\gamma$. 
The symmetrical feature with respect to $\gamma=1/g=1$ given by 
eq.~(\ref{eq:symGWF}) is observed for half filling in both components. 
As $n$ decreases from 1, this symmetry is deservedly broken. 
Henceforth, we will often use $\gamma$ instead of $g$, because
$\gamma$ in AHM plays a substantially equivalent role to $g$ in RHM. 
\par

\begin{figure}
\epsfxsize=6cm
\epsfysize=4cm
\caption{
Expectation values of (a) ${\cal H}_t$ and (b) ${\cal H}_U$ with 
respect to GWF eq.~(\ref{eq:GWF}) are shown for several electron 
densities. 
The data for $\gamma\le1$ ($\gamma\ge 1$) correspond to the attractive 
(repulsive) case. 
Samples as many as $5\times 10^5$-$5\times 10^6$ are used for each data 
point in VMC calculations. 
Statistical errors are negligible in this scale. 
}
\label{fig:EsymGWF}
\end{figure}

\subsection{Results of energy}
Now, we reconsider the energy of GWF by VMC calculations. 
In Fig.~\ref{fig:EGWF} expectation values of kinetic and potential 
energies are plotted versus $\gamma$ in the attractive regime. 
Since onsite pairs are more favored as $\gamma$ decreases from 1, 
the ratio of double occupation $d$ $(=E_U/U)$ increases, and the system 
loses kinetic energy. 
In the limit of $\gamma\rightarrow 0$, all the electrons form 
onsite pairs for arbitrary electron density: $d=n/2$. 
Thus, $E_t$ vanishes for any value of $n$. 
This feature is in sharp contrast to the repulsive case, in which 
$E_t$ has finite transfer even at $g=0$ except for $n=1$. 
Since the behavior of $E_t/t$ and $E_U/U$ in this limit is essential 
for the Brinkman-Rice metal insulator transition (BRT), we will 
investigate their behavior in detail in Appendix A. 
\par

\begin{figure}
\epsfxsize=6cm
\epsfysize=2cm
\caption{
Expectation values of (a) ${\cal H}_t$ and (b) ${\cal H}_U$ by GWF 
for all the possible electron densities for $L=10$ in the attractive 
regime. 
$5\times 10^5$-$10^7$ samples are used for each value of $\gamma$ 
in VMC calculations. 
Statistical errors are by far smaller than dots. 
}
\label{fig:EGWF}
\end{figure}

Total energy $E$, the sum of $E_t$ and $E_U$, has a minimum at 
$0<\gamma<1$ for any finite value of negative $U$, because BRT never 
arises in finite dimensions as discussed in Appendix A. 
As an example Fig.~\ref{fig:TEGWF} shows $E/t$ for two values of $U/t$ 
at half filling; the values for $L=\infty$ (cross) are estimated 
by some extrapolations.\cite{noteSSD} 
With increasing $|U|/t$ the optimal value of $\gamma$ naturally decreases. 
Here, note the difference between Figs.~\ref{fig:TEGWF}(a) and 
(b) that the optimal value of $\gamma$ hardly depends on $L$ for $U/t=-8$, 
while it remarkably decreases with increasing $L$ for $U/t=-16$. 
By checking the data for various values of $U/t$, it is found that 
the former (latter) feature is common to the regime of small (large) 
$|U|/t$, and a rather precipitous crossover takes place at  
$U\equiv U_{\rm co}\sim -11t$. 
\par

\begin{figure}
\epsfxsize=6cm
\epsfysize=2cm
\caption{
Total energy of GWF at (a) $U/t=-8$ and (b) $-16$ for several system 
sizes at half filling. 
Symbols are common in (a) and (b). 
Minimal points are indicated by solid circles connected by dashed lines. 
The crosses on these lines indicate the values in the thermodynamic limit,
which are extrapolated from the data of every even $L$ ($6\le L\le 20$) 
and also $L=24$ for $U/t=-16$ by second-order polynomial fit of $1/L^2$. 
The value of GA is also shown in (a); as for (b), the total energy of 
GA is $U/2=-8t$ because $|U|/t>|U_{\rm BR}|/t=12.969\cdots$. 
Incidentally, although $E/t$ of GWF is always lower than the value 
of GA in Fig.~\ref{fig:TEGWF}, this is not always true, because GA 
violates the variation principle.\cite{YS1} 
}
\label{fig:TEGWF}
\end{figure}

Such a crossover is recognizable also in Fig.~\ref{fig:optGWF}(a), 
where the optimized values of $\gamma$, $E_t/t$ and $E_U/U$ are 
shown for a couple of system sizes. 
With increasing $|U|/t$, $\gamma$ at first shows a rapid decrease 
almost without system-size dependence. 
When $U$ reaches $U_{\rm co}$, however, the decrease of $\gamma$ 
becomes slow and asymptotic with respect to $\gamma=0$, and 
simultaneously system-size dependence becomes conspicuous. 
With this change of $\gamma$, $|E_t|/t$ decreases with weak system-size 
dependence originating in the noninteracting band (not by $\gamma$), 
whereas large dependence shows up for $|U|\gsim|U_{\rm co}|$ due to
the behavior of $\gamma$. 
As for the potential part, double occupancy $d$ increases and almost 
saturates for $|U|\gsim|U_{\rm co}|$, and its system-size dependence 
is reversed at $U\sim U_{\rm co}$. 
\par

\begin{figure}
\epsfxsize=6cm
\epsfysize=2cm
\caption{
Optimized values of $\gamma$ and the energy components $E_t/t$ and 
$E_U/U$ as a function of $|U|/t$ at (a) half filling and 
(b) a quarter-filling. 
For clarity only a couple of system sizes are taken up. 
The results of GA are also plotted, in which BRT takes place 
at $U_{\rm BR}$. 
}
\label{fig:optGWF}
\end{figure}

Plotted in Fig.~\ref{fig:totEGWF} is the total energy $E/t$ 
measured from that of the completely localized or paired state: 
$E_\infty=Un/2$.
The magnitude of this quantity rapidly reduced with increasing $|U|/t$ 
for small $|U|$; around $U=U_{\rm co}$, however, the behavior is 
switched to $-t/|U|$. 
Thus, GWF represents two kinds of normal states, according as 
$|U|\lsim|U_{\rm co}|$ or $|U|\gsim|U_{\rm co}|$. 
In the former region the state obviously belongs to a Fermi liquid, 
which continues to the noninteracting case, whereas in the latter 
most of the electrons form onsite pairs as in Fig.~\ref{fig:optGWF}(a), 
and the state looks consistent with the picture of hard-core bosons 
represented by eq.~(\ref{eq:ehamil}), in which the kinetic contribution 
does not vanish but is proportional to $t^2/U$. 
As will be seen in the next subsection, however, this state is still 
metallic---an almost localized Fermi liquid---, because a clear Fermi 
surface is defined and there is no excitation gap. 
This is in contrast with the GA [solid line in Fig.~\ref{fig:optGWF}(a)], 
which brings about a metal-insulator transition at $U=U_{\rm BR}$. 
In the state for $|U|>|U_{\rm BR}|$ all the electrons stand still and 
$E_t$ vanishes, although it is quantitatively good for $|U|<|U_{\rm BR}|$. 
This unfavorable feature is the same with the insulating state obtained 
by DMFT.\cite{MetznerDMF,Capone}
\par

\begin{figure}
\epsfxsize=6cm
\epsfysize=2cm
\caption{
Total energy of GWF at half filling as a function of $|U|/t$ measured 
from $E_\infty=nU/2$, namely the energy of the completely paired state. 
Shown is the value for $L=\infty$ extrapolated from the data of $L=6$-20 
and 24 (for large values of $|U|/t$) as in Fig.~\ref{fig:TEGWF}. 
The thin dashed line with an arrow shows the $-t/|U|$ curve whose 
coefficient is chosen so as to fit the data of $L=\infty$.
The inset shows the magnification of small-$E$ region; 
the data for some finite $L$'s are included. 
Incidentally, this quantity is equivalent to the simple total energy 
of RHM at half filling [See eq.~(\ref{eq:symGWF})]. 
}
\label{fig:totEGWF}
\end{figure}

Now, we turn to less-than-half filling.
Plotted in Fig.~\ref{fig:optGWF}(b) is the same quantities with 
Fig.~\ref{fig:optGWF}(a) but for a quarter filling. 
As compared with half filling, the qualitative feature scarcely 
changes including system-size dependence. 
Therefore, fixing the system size, we focus on the $n$ dependence.
In Fig.~\ref{fig:optGWFn} we show the optimal values of the energy 
components versus $|U|/t$ for every available value of $n$ for $L=10$. 
Both $E_t$ and $E_U$ change their behavior around $U=U_{\rm co}$ 
from precipitous increase with increasing $|U|/t$ ($U\lsim U_{\rm co}$) 
to almost saturated one ($U\gsim U_{\rm co}$). 
As has repeatedly insisted, in the former region the kinetic term is 
dominant (Fermi liquid), while in the latter region the potential 
term is superior (almost localized). 
An important point here is that although the absolute values of energies 
are directly related to the electron density, the values of $|U_{\rm co}|$ 
once slightly increases with decreasing $n$, has a broad maximum near 
$n=0.5$ and shifts to a somewhat smaller value in the vicinity of $n=0$. 
On the whole this behavior agrees with that of $|U_{\rm BR}|$ in 
GA (Fig.~\ref{fig:BRT}), although $|U_{\rm co}|$ is smaller than 
$|U_{\rm BR}|$ to some extent. 
To summarize, in contrast to RHM, an almost localized state exists 
also for less-than-half filling in a similar range of $U/t$ to 
half filling. 
\par

\begin{figure}
\epsfxsize=6cm
\epsfysize=2cm
\caption{
Energy components (a) $E_t/t$ and (b) $E_U/U\ (=d)$ at the optimal 
$\gamma$ as a function of $|U|/t$ for all the electron densities 
available for $L=10$. 
The results of $\Psi_Q$ ($L=10$, $n=1.0$) treated in \S4 are also 
shown for comparison by shaded dashed lines. 
Symbols are common to both figures. 
Energy minimization is carried out based on the data points with 
$10^6$-$5\times 10^6$ samples. 
}
\label{fig:optGWFn}
\end{figure}

\subsection{Results of correlation functions}

In this subsection we study GWF through momentum distribution, 
$ 
n(k)=\frac{1}{2}\sum _\sigma 
     \langle c_{k\sigma}^\dagger c_{k\sigma}\rangle, 
$
and a couple of correlation functions defined as,
\begin{eqnarray}
S^z(q)&=&\frac{4}{N}\sum_{j\ell}e^{-iqr_\ell}\langle S^z_jS^z_{j+\ell}\rangle, 
\qquad\quad {\rm (spin}\ z\ {\rm component)}
\\
\label{eq:nksq}
N(q)&=&\frac{1}{N}\sum_{j\ell}e^{-iqr_\ell}\langle n_jn_{j+\ell}\rangle
-n^2,
\qquad {\rm (charge\ density)}
\\
P(q)&=&\frac{1}{N}\sum_{j\ell}e^{-iqr_\ell}
\langle \Delta_j^\dagger\Delta_{j+\ell}\rangle,
\qquad\quad {\rm (onsite\ pairing)}
\end{eqnarray}
where $n_j=n_{j\uparrow}+n_{j\downarrow}$ and
$\Delta_j^\dagger=c_{j\uparrow}^\dagger c_{j\downarrow}^\dagger$. 
Some properties of these quantities have been already studied with 
$\Psi_{\rm G}$ for the 1D and 2D RHM.\cite{YS1,YS3} 
Although they partly exhibit unphysical behavior,\cite{noteGWFcor} 
they are suggestive and useful to compare AHM with RHM. 

To apply the previous results of RHM to AHM, we summarize the relationship 
among correlation functions through the canonical transformation 
eq.~(\ref{eq:canon}). 
Since $N_\ell=\langle n_jn_{j+\ell}\rangle
-\langle n_j\rangle\langle n_{j+\ell}\rangle$ is transformed to 
$\tilde S^z_\ell=4(\langle\tilde S^z_j\tilde S^z_{j+\ell}\rangle
-\langle\tilde S^z_j\rangle\langle\tilde S^z_{j+\ell}\rangle$), 
$N(q)$ in AHM corresponds to $S^z(q)$ in RHM, as 
anticipated from the relation between $O_{\rm CDW}$ and $O_{\rm SDW}^z$. 
Similarly, $P(q)$ in AHM is related to the staggered spin correlation 
in RHM as, $P(q)=\tilde S^\pm(q+G)/2$, because 
$\Delta_j^\dagger\Delta_{j+\ell}$ is transformed to 
$(-1)^j\tilde S^+_j\tilde S^-_{j+\ell}$, where 
\begin{equation}
S^\pm(q)=\frac{1}{N}\sum_{j\ell}e^{-iqr_\ell}
\langle S^+_jS^-_{j+\ell}+S^-_jS^+_{j+\ell}\rangle. 
\label{eq:sqpm}
\end{equation}
Using these relations and assuming 
$\tilde S(q)=\tilde S^z(q)=\tilde S^\pm(q)$, we have $N(q)=2P(q-G)$. 
In particular, the case for $q=G$ [$N(G)=2P(0)$] tells that the 
correlation of antiferro CDW is always proportional to the correlation 
of onsite singlet pairing, if the wave function, which may be exact 
or approximate, remains isotropic in spin space after the transformation. 
Since this condition is satisfied by GWF at half filling due to 
eqs.~(\ref{eq:bFermisea}) and (\ref{eq:tildeGWF}), the relation 
$[S(q)]_g=[N(q)]_\gamma=2[P(q-G)]_\gamma$ holds, where $[\cdots]_g$ 
indicates the expectation value with respect to $\Psi_{\rm G}(g)$. 
However, this is not the case for $n\ne 1$, where $\Psi_{\rm G}$ does not 
meet the condition for spin isotropy. 
As for momentum distribution, particularly at half filling one can 
confirm the relation $[n(k)]_g=[n(k)]_\gamma$ by using the inversion 
and electron-hole symmetries in addition to the transformation 
eq.~(\ref{eq:canon}).
\par

To begin with, we look at the half-filled case for $U<0$, first. 
In Figs.~\ref{fig:GWFcf}(a) we show $n(k)$ for several attractive 
values of $g$, which are equivalent to the repulsive case with $1/g$ 
or with $|U|/t$. 
Note that the discontinuity at $k=k_{\rm F}$ is clear for 
$|U|>|U_{\rm co}|$, not to mention $|U|<|U_{\rm co}|$. 
In Fig.~\ref{fig:renorm} we show the renormalization factor $Z$ 
(quasi particle weight in the Fermi liquid theory) estimated from 
the discontinuity of $n(k)$ at $k=k_{\rm F}$: 
$Z=n(k_{\rm F}-0)-n(k_{\rm F}+0)$; 
$Z$ of GWF decreases similarly to GA for small values of $|U|/t$, 
but abruptly changes the behavior around $U=U_{\rm co}\sim -11t$ 
to an asymptotic one. 
Thus, we confirm GWF remains metallic also in the strong-coupling 
side of $U_{\rm co}$, even though the effective mass of electrons 
($1/Z$) is considerably large. 
\par

Next, we touch on correlation functions. 
According to the previous studies for $U>0$, $N(G)$ and $P(0)$ should 
be enhanced and $S(q)$ is suppressed by attractive correlation; this 
aspect is actually seen in Fig.~\ref{fig:GWFcf}(b), where $N(G)$ becomes 
overwhelmingly large. 
Therefore, we check the possibility of the divergence of $N(G)$ as well 
as the realization of a long-range order in Appendix B. 
Here, we focus on the gaps in excitation spectra, which we learn from 
the behavior of $N(q)$ and $S(q)$ near the ${\rm\Gamma}$ point 
($|q|\rightarrow 0$). 
In this limit of $|q|$, both $N(q)$ [inset of Fig.~\ref{fig:GWFcf}(b)] 
and $S(q)$ [$N(q)$ in Fig.~1(a) of ref.\citen{YS3}] behave as a linear 
function of $|q|$ irrespective of the values of $U/t$. 
Therefore, GWF is always gapless both in spin and charge degrees of 
freedom, supporting GWF's metallic properties in both sides of $U_{\rm co}$. 
This aspect is preserved in less-than-half filling as observed in 
Figs.~\ref{fig:cfqf}(a) and (d). 
\par

\begin{figure}
\epsfysize=2cm
\caption{
(a) Momentum distribution function and (b) charge density structure factor 
of GWF at half filling for several attractive values of $g$; 
the corresponding values of $U/t$ [common to (a)] are shown in (b). 
The values for $|U|<|U_{\rm co}|$ ($|U|>|U_{\rm co}|$) are shown by solid 
(open) symbols. 
Inset in (b) shows the magnification of $N(q)$ on the ${\rm \Gamma}$-X line. 
In (a) each $k$ point shifts by $\pi/L$ in $y$ direction due to the 
antiperiodic boundary condition; a mark with a prime (\eg ${\rm \Gamma}'$) 
indicates a shifted point. 
The arrow shows the position of ${\rm M}'$ point. 
}
\label{fig:GWFcf}
\end{figure}
\par

Now, we turn to the less-than-half filling, where the relation to 
RHM is not so simple as at half filling, and $N(G)$ and $2P(0)$ are 
no longer equivalent. 
Since the correlations of charge density and onsite singlet pairing are 
dominant in AHM, we take up $N(q)$ and $P(q)$, first. 
To grasp the $n$ dependence we fix the parameters at $U/t=-8$ and 
$L=10$ (Fig.~\ref{fig:cflthf}). 
With decreasing $n$ the sharp peak of $N(q)$ at $q=G$ is rapidly 
suppressed, because the nesting by $G$ is lost. 
Instead, weak singular behavior appears at incommensurate wave numbers 
[$q=2k_{\rm F}$ or $2(G-k_{\rm F})]$ of the noninteracting case for 
$n\lsim 0.72$.\cite{noteincomme} 
When $n$ further decreases, $N(q)$ tends to be small and flat, namely 
charge correlation becomes weak and restricted to a considerably 
short range. 
We show $|U|/t$ dependence of $N(G)$ for $L=10$ in Fig.~\ref{fig:cfnvsu}(a). 
At half filling, $N(G)$ shows precipitous increase around $U=U_{\rm co}$, 
but slowly saturates for $|U|\gsim|U_{\rm co}|$. 
When $n$ decreases and $G$ (M point) is no longer a special $q$-point, 
the conspicuous change around $U\sim U_{\rm co}$ vanishes. 
\par

In contrast to $N(q)$, $P(q)$ preserves the peak at $q=(0,0)$ even for 
the lowest electron density $n=0.04$ as shown in Fig.\ref{fig:cflthf}(b), 
although its magnitude gradually diminishes with $n$. 
Thus, the singlet pairing correlation remains dominant for all $n$ in GWF, 
although it does not develop into a long-range order (see Appendix B). 
As for $|U|/t$ dependence [Fig.~\ref{fig:cfnvsu}(b)], $P(0)$ abruptly 
increases around $U=U_{\rm co}$ for every electron density. 
Note that for $|U|>|U_{\rm co}|$ the absolute values of $P(0)$ scarcely 
changes with $n$ for $n\gsim 0.72$, and has rather larger values for 
$n=0.96$ and 0.88 than for half filling. 
Concerning $S(q)$, which is considerably suppressed and flattened by $U$ 
already at half filling, the tendency toward shortened correlation 
is further promoted by decreasing $n$ [inset of Fig.~\ref{fig:cflthf}(b)]. 
\par

\begin{figure}
\epsfxsize=6cm
\epsfysize=2cm
\caption{
Structure factors of (a) charge density and (b) onsite singlet pairing 
and spin (inset), for all the electron densities available in $L=10$. 
Symbols are common to all the panels. 
In each panel the value for $U/t=0$ ($L=\infty$) at half filling is 
shown by a dashed line. 
We take a path ${\rm \Gamma}\rightarrow {\rm X}\rightarrow {\rm M}
\rightarrow {\rm \Gamma}$ in a Brillouin zone. 
$5\times 10^5(n=1.0)$-$5\times 10^6 (n=0.04)$ samples are used. 
}
\label{fig:cflthf}

\end{figure}
\begin{figure}
\epsfxsize=6cm
\epsfysize=2cm
\caption{
Structure factor of (a) charge density at $q=G$ and (b) onsite single 
pairing at $q=0$ as a function $|U|/t$ for all the electron densities 
available in $L=10$.
}
\label{fig:cfnvsu}
\end{figure}

Finally, we consider the difference between AHM and RHM for $n<1$. 
Shown in Figs.~\ref{fig:cfqf}(c) and (d) are $n(k)$ and $S(q)$ of GWF 
at a quarter filling for several attractive values of $g$, respectively. 
One finds similar features to those at half filling, for instance, 
[1] $n(k)$ is an increasing function of $|k|$, and 
[2] has clear discontinuities at $k=k_{\rm F}$ for any finite 
value of $g$. 
[3] $Z$ changes its behavior at $U\sim U_{\rm co}$ (See 
Fig.~\ref{fig:renorm}). 
[4] $S(q)$ is remarkably suppressed with increasing correlation strength 
$|U|/t$. 
\par

As for [1], the increase of $n(k)$ especially inside the Fermi surface 
in Fig.\ref{fig:cfqf}(c) is not only common to but more remarkable 
than that at half filling [Fig.\ref{fig:GWFcf}(a)]. 
This tendency is contrary to that in less-than-half filling of RHM 
[cf. Fig.~2(b) of ref.~\citen{YS3}].\cite{noteGWFcor}
A similar situation, that is, the feature is analogous between $n<1$ 
and $n=1$, is also observed in $N(q)$ [Fig.~\ref{fig:cfqf}(a)] and 
$S(q)$ for AHM.
As mentioned, $N(q)$ comes to have higher peaks (though not so dominant 
as for $n=1$ [Fig.~\ref{fig:GWFcf}(b)]) at 
$k=2k_{\rm F}$ and $2(G-k_{\rm F})$ with increasing $g$. 
Meanwhile, $S(q)$ in less-than-half filling for RHM does not have peaks 
at those wave numbers but only cusps [See Fig.~2(a) of ref.~\citen{YS3}].
As for $S(q)$---the point [4] [Fig.~\ref{fig:cfqf}(d)]---, the maximum 
for $g-1=50\ (U/t\sim -15)$ is no more than 1.6\% of the maximum value 
0.5 for $g-1=0\ (U/t=0)$; the corresponding value at half filling 
[$g-1=50\ (U/t\sim -15),\ L=20$] is 0.85\%. 
In both densities, the reduction of $S(q)$ is considerable. 
On the other hand, the corresponding quantity [$N(q)$ for $U/t=16$, 
$L=16$] at a quarter filling of RHM preserves its amplitude as large as 
more than 60\% of the free case [Fig.~2(a) of ref.~\citen{YS3}]. 
\par

Accordingly, the state for $n<1$ in AHM has analogous properties of 
correlation functions to its counterpart in RHM at half filling, but 
not for $n<1$ in RHM at least within GWF. 
\par

\begin{figure}
\epsfxsize=6cm
\epsfysize=2cm
\caption{
Structure factors of (a) charge density, (b) onsite singlet pairing 
and (d) spin, and (c) momentum distribution by GWF at a quarter filling 
for several values of attractive $g$ ($\gamma$), which correspond 
to the values of $U/t$ shown in (b) and (c). 
Arrows on the noninteracting case ($g=1$) indicate the points of 
$2k_{\rm F}$ or $2(G-k_{\rm F})$. 
The corresponding singular behavior in $N(q)$ is indicated by the 
letters A, B and C in (a). 
Path in $q$ space is the same with Figs.~\ref{fig:cflthf}
and \ref{fig:GWFcf}(a).
$2.5\times10^5$ samples are used for each parameter set. 
}
\label{fig:cfqf}
\end{figure}

\section{Intersite correlation factor} 

As discussed in our previous papers\cite{YS1,YS3} GWF is too simplistic 
to reproduce properly some aspects of momentum distribution and 
correlation functions.\cite{noteGWFcor} 
This is primarily because intersite correlation factors are neglected 
in GWF, as we showed for RHM.
In this section and Appendix C we improve the normal-state wave function 
on GWF by taking account of these factors. 
In Appendix C we take up a distance-dependent Jastrow-type correlation 
factor, which was successful for less-than-half filling of RHM.\cite{YS3} 
On the other hand, in this section a binding effect between up and down 
spins in adjacent sites is introduced by allowing for the virtual processes 
in the perturbation expansion in the strong coupling limit. 
In \S4.1 we introduce the binding wave function $\Psi_{\rm Q}$. 
In \S4.2 and \S4.3 the results at half filling and in less-than-half filling 
are presented, respectively. 
Section 4.4 is assigned to discussions and comparisons with other results. 
A further discussion on the metal-insulator transition in and beyond 
$\Psi_Q$ is given in Appendix D. 
\par 

\subsection{A binding factor between up- and down-spin sites}
Consider the strong coupling limit ($t/|U|\rightarrow 0$). 
All the electrons form onsite pairs in the ground state 
[Fig.~\ref{fig:schfig}(a)]. 
A small transfer term causes an on-site pair broken, yielding thereby 
virtually excited $\uparrow$- and $\downarrow$-spin pairs next to 
each other [Fig.~\ref{fig:schfig}(b)]; they corresponds to the virtual 
states in the lowest-order perturbation processes. 
Such a configuration should have a larger weight than ones for 
higher-order processes, which include isolated singly-occupied 
sites as in Fig.~\ref{fig:schfig}(c). 
In fact, improvement in this line has been already embodied for 
the half-filled RHM,\cite{KHF-Fazekas,YS3} where the binding between 
doubly-occupied (d-) and empty (e-) sites is vital to the virtual 
processes. 
The wave function thus introduced ($\tilde\Psi_Q$)\cite{YS3}
was extremely good, especially for the 1D half-filled band, 
not only in energy but for various correlation functions. 
\par

\begin{figure}
\epsfxsize=6cm
\epsfysize=2cm
\caption{
Schematic examples to assign weights of the correlation factor in 
$\Psi_{\rm Q}$ on 2D square lattice. 
Each figure represents a local electron configuration, where an open 
circle indicates an empty site, a solid circle a doubly-occupied one 
and an upward (a downward) arrow a singly occupied one with an up 
(a down) spin. 
(a) A configuration realized in the ground state at $U/t=-\infty$. 
Here, we assume the weight of this configuration as 1.
(b) A configuration in which one electron pair [marked with a cross in (a)] 
is broken by a one-step hopping process (a virtual state in the 
second-order perturbation processes). 
The weight $\gamma(=1/g)$ is assigned due to the Gutzwiller factor. 
(c) A configuration in which one electron pair is broken and two hopping 
processes are needed to reach from (a) (a virtual state of the 
fourth-order perturbation processes). 
The weight is suppressed by the factor $(1-\mu)^2$ in addition to $\gamma$. 
}
\label{fig:schfig}
\end{figure}

For the attractive case a suitable wave function is expected by applying 
the canonical transformation eq.~(\ref{eq:canon}) to $\tilde\Psi_{\rm Q}$, 
and consequently we have, 
\begin{equation}
\Psi_Q=\prod_j\left[1-\mu Q_j\right]\Psi_{\rm G}(\gamma),
\label{eq:udbinding}
\end{equation} 
\begin{equation}
Q_j=s_j^\uparrow\prod_\tau(1-s_{j+\tau}^\downarrow) 
    +s_j^\downarrow\prod_\tau(1-s_{j+\tau}^\uparrow), 
\label{eq:udbindingf}
\end{equation}
where $s_j^\sigma=n_{j\sigma}(1-n_{j-\sigma})$ and $\tau$ runs over 
all the nearest neighbors. 
In $\Psi_Q$ the second parameter $\mu$ ($0\le\mu\le 1$) controls 
the attractive correlation between a pair of up- and down-spin 
electrons in next-nearest-neighbor sites both of which are singly 
occupied as explained in Fig.~\ref{fig:schfig}. 
When $\mu=0$, $\Psi_{\rm Q}$ is reduced to GWF. 
As $\mu$ increases, singly-occupied sites without an antiparallel-spin 
site in nearest neighbors become disadvantageous. 
The state becomes insulating at last for $\mu=1$, because then an 
up-spin and a down-spin electron are completely bound within 
nearest-neighbor sites, unless they are on the same site. 
A merit of $\Psi_Q$ among similar functions\cite{noteFazekas} is 
that the on-site correlation and the nearest-neighbor binding 
can be treated independently by $g$ and $\mu$, respectively. 
\par

In particular, at half filling the relation eq.~(\ref{eq:symGWF}) holds 
also between $\tilde\Psi_{\rm Q}(g,\mu)$ and $\Psi_{\rm Q}(\gamma,\mu)$, 
because the correlation factor for binding in eq.~(\ref{eq:udbinding}) 
is invariant under the canonical transformation and GWF satisfies 
eq.~(\ref{eq:tildeGWF}).
Thus, the results obtained in RHM\cite{YS3} can be directly applied 
to AHM at half filling. 
For example, $\tilde\Psi_Q$ is lower in energy than an antiferromagnetic 
state ($\tilde\Psi_{\rm SDW}$)\cite{YS2} in 1D, but higher in 2D for 
RHM.\cite{YS3} 
This reads for AHM as $\Psi_Q$ is stabler than $\Psi_{\rm CDW}$\cite{noteCDW} 
in 1D, but not in 2D. 
\par 

\subsection{Results for half-filled band}
First of all, let us reconsider the case of half filling. 
In Fig.~\ref{fig:udetu} expectation values of the energy components 
are shown for various values of the parameters. 
As for the kinetic part, $E_t$ once slightly decreases with $\mu$ and have 
a minimum at a small value of $\mu$ for $\gamma<1$ 
[inset of Fig.~\ref{fig:udetu}(a)]. 
This decrease of $E_t$ causes the gain in total energy for small $|U|$. 
For $\mu\gsim 0.2$, $E_t$ comes to increase monotonically, but remains 
finite even for $\mu=1$ due to the second-order perturbation processes. 
On the other hand, double occupancy $E_U/U$ decreases with increasing 
$\mu$ for $\gamma\gsim 0.5$ due to the fluctuation effect of $\mu$, 
whereas $d$ increases with $\mu$ for smaller values of $\gamma$
[inset of Fig.~\ref{fig:udetu}(b)], implying that isolated unpaired 
electrons are confined in local singlet pairs. 
Anyway, $\Psi_{\rm Q}$ has lower energy than GWF for any negative 
value of $U$, as we showed in the previous study for RHM.
The energy decrement is remarkable especially in the intermediate 
correlation regime [See Fig.~\ref{fig:udDE}]. 
In Fig.~\ref{fig:udetot} total energy for various parameter values 
are depicted as an example for large-$|U|$ region. 
\par

\begin{figure}
\epsfxsize=6cm
\epsfysize=2cm
\caption{
Expectation values of (a) kinetic and (b) potential energies of 
$\Psi_{\rm Q}$ as a function of $\gamma=1/g$ at half filling. 
Insets show $\mu$ dependence for respective quantities.  
Symbols are common in (a) and (b).
Although the system with $L=6$ is plotted for abundant data points, 
it is confirmed that the system-size dependence is quantitatively 
small. 
We use $10^6$ samples for each data point. 
Statistical errors are much smaller than the symbol size. 
}
\label{fig:udetu}
\end{figure}

\begin{figure}
\epsfxsize=6cm
\epsfysize=2cm
\caption{
Example of energy expectation values of $\Psi_{\rm Q}$ for a large 
value of $|U|/t$ as a function of binding correlation strength $\mu$ 
for several values of onsite attractive correlation $g$. 
The arrow indicates the minimum -8.2118(23), which is given by $\mu=0.9$ 
and $1-g=-7.0$ and much lower than the value of GWF ($\mu=0$) -8.0463(12)
but comparable to the insulating ($\mu=1$) value -8.2115(23).
}
\label{fig:udetot}
\end{figure}

Next, we look at the optimized variational parameters 
(Fig.~\ref{fig:udparam}). 
For small $|U|(\lsim W)$ the optimized value of $\mu$ greatly 
increases with increasing $|U|/t$, and its system-size dependence 
is small. 
For $|U|\gsim W$, however, $\mu$ comes to change slowly near 1, 
and becomes highly dependent on $L$ and $n$. 
On the other hand, the optimal value of $\gamma$ for $\Psi_{\rm Q}$ 
is almost the same with (slightly smaller than) that of GWF for 
$|U|\lsim W$, undergoes a cusp-like change at $U\sim W$ 
and becomes considerably larger for $|U|\gsim W$. 
The dependence on $L$ and $n$ is weak for any value of $U$. 
These facts indicate that there is some crossover at $|U|=|U_Q|\sim W$ 
and the state in the strong-coupling side is qualitatively different 
from GWF, and nearly insulating. 
\par

\begin{figure}
\epsfxsize=6cm
\epsfysize=2cm
\caption{
Optimized values of the parameters in $\Psi_Q$ versus $|U|/t$ for some 
system sizes and densities. 
Conspicuous errors in $\mu$ for $|U|\gsim|U_Q|$ stem from the flat 
bottom of $E/t$ near $\mu=1$ (See \S4.4). 
For comparison, the optimal values of $\gamma$ for GWF are also shown. 
}
\label{fig:udparam}
\end{figure}

To study this crossover closely we consider energy components $E_t/t$ 
and $d$, which are shown in Figs.~\ref{fig:udeopt}(a) and (b), respectively. 
With increasing $|U|/t$, $|E_t|$ abruptly decreases until $|U|$ reaches 
$|U_Q|$, around which $|E_t|$ changes its behavior, and comes to decrease 
slowly for $|U|>|U_Q|$. 
Note that as $L$ increases, the change of behavior at $U\sim U_Q$ 
becomes more abrupt and cusp-like. 
A similar abrupt change at $U\sim U_Q$ can be seen in $E_U$. 
Thus, an abrupt crossover or a phase transition\cite{notetrans} takes 
place around $U=U_Q$, namely $\Psi_{\rm Q}$ represents two distinct 
normal (or incoherent) states. 
The critical value $U_Q$ is broadly estimated as $-8.8$ at half filling 
by finding the intersection of a smoothly extrapolated curve from 
$|U|<|U_Q|$ and a line of $\mu=1$ with respect to $d$ 
[Fig.~\ref{fig:udeopt}(b)], which we adopt on account of its sharper 
change there. 
The value thus obtained does not largely depend on $L$. 
As will be explained below with correlation functions, in the region of 
$|U|<|U_Q|$ the state is a Fermi liquid, which is smoothly 
connected to the noninteracting case, whereas the state for 
$|U|>|U_Q|$ is very akin to the insulating state ($\mu=1$) 
indicated by dashed lines in Figs.~\ref{fig:udeopt}. 
In this nearly insulating state almost all the electrons are paired 
either at the same site or within the nearest-neighbor sites. 
As compared with the optimal GWF (Fig.\ref{fig:optGWFn}), double 
occupancy in $\Psi_{\rm Q}$ has a larger value around $U=U_Q$ 
but is only a little smaller for $|U|\gsim|U_{\rm co}|(\sim 11t)$ 
due to the fluctuation effect of $\mu$.
Meanwhile $\Psi_{\rm Q}$ is considerably advantageous in kinetic 
energy in the strong coupling side, thanks to the second-order 
perturbation term. 
\par

\begin{figure}
\epsfxsize=6cm
\epsfysize=2cm
\caption{
Expectation values of (a) kinetic and (b) potential energies by the 
optimized parameter sets as a function of $|U|/t$ at half filling. 
Four system sizes are shown. 
Dashed lines denote the optimized insulating values ($\mu=1$). 
Insets show the same quantities for four values of electron density
with a fixed system size $L=10$. 
We have used $5\times 10^5$-$2\times 10^6$ samples for energy 
minimization (each parameter set). 
}
\label{fig:udeopt}
\end{figure}

Now, we look at this transition in momentum distribution function 
shown in Fig.\ref{fig:udnkhf} for various values of $U$. 
In the region of $|U|<|U_Q|$ discontinuities at the two $k_{\rm F}$ 
points are easily recognized, indicating a degenerate metallic state. 
On the other hand, for $|U|\gsim|U_Q|$ the discontinuity at 
$k_{\rm F}$ vanishes or is at least not obvious, so that the state 
is insulating. 
Furthermore, near this boundary there is a qualitative distinction in 
system-size dependence between the two region (not shown), namely 
the discontinuity at $k_{\rm F}$ increases with increasing $L$ for 
$|U|<|U_Q|$ (\eg $U=-8t$), whereas it rapidly decreases and 
tends to vanish for $|U|>|U_Q|$ (\eg $U=-9t$). 
In Fig.\ref{fig:renorm} renormalization factor $Z$ is depicted (solid 
circle) as a function of $|U|/t$. 
$Z$ comes to decrease remarkably as $U$ approaches $U_Q$ and 
vanishes there; the effective mass of quasi particles 
diverges and the state becomes insulating. 
Thus, the existence of a metal-insulator transition is ascertained
(See Appendix D for details). 
\par

\begin{figure}
\epsfxsize=6cm
\epsfysize=2cm
\caption{
Momentum distribution function by $\Psi_{\rm Q}$ at half filling 
for various values of $U/t$. 
Closed (open) symbols are used for the Fermi liquid (insulating) 
regime; $U/t=-9$ is near the critical point. 
The path in $q$ space is the same with Fig.~\ref{fig:GWFcf}(a). 
We use $10^6$ samples for each optimized parameter set. 
}
\label{fig:udnkhf}
\end{figure}

\begin{figure}
\epsfxsize=6cm
\epsfysize=2cm
\caption{
Renormalization factor (quasi particle weight) in Fermi liquid states 
versus $|U|/t$ estimated from the discontinuity of $n(k)$ at $k=k_{\rm F}$ 
on the ${\rm \Gamma}$-X line for some electron densities. 
Results of GWF and GA in 2D and of DMFT for hypercubic lattice\cite{Bulla} 
are also included. 
}
\label{fig:renorm}
\end{figure}

Let us move on to $N(q)$,\cite{noteudnq} which has a characteristic 
peak of half filling at the M point as seen in Fig.\ref{fig:udnqhf}.
In the inset $N(G)$ is plotted versus $U/t$; the increase of $N(G)$ with 
$|U|/t$ is abrupt right below $|U|=|U_Q|$, but becomes slow for $|U|>|U_Q|$. 
Thus, the effect of transition is clear also in $N(q)$. 
At these sizes the magnitude of $N(G)$ for large $|U|$ is about an 
order of magnitude smaller than that of $\Psi_{\rm J}$ in the CDW phase 
(Fig.~\ref{fig:JNG}), is not proportional to system size 
even though increasing with $L$, and is the same degree with that 
of GWF (Fig.~\ref{fig:cfnvsu}). 
In addition, $O_{\rm CDW}$ is always zero irrespective of the value of 
$L$ within the statistical error. 
It follows that only short-range correlations grow even for 
$|U|\gsim|U_Q|$ in $\Psi_Q$. 
\par

Finally, we look at excitation gaps. 
As for the charge degree of freedom, $N(q)\sim \nu|q|$ for 
$|q|\rightarrow 0$ as seen in Fig.~\ref{fig:udnqhf}, and $\nu$ only 
weakly depends on the value of $|U|/t$. 
Thus, there is no charge gap in both phases. 
On the other hand, there is a distinction in the small-$|q|$ behavior 
of $S(q)$ between the metallic and the insulating phases, as observed 
in the inset of Fig.~\ref{fig:udcfpf}(b). 
For $|U|<|U_Q|$ the leading term of $S(q)$ seems linear (or a smaller 
exponent), while for $|U|>|U_Q|$ it seems $S(q)\propto |q|^\alpha$ 
with $\alpha>1$. 
Thus, we conclude that a spin gap does not exist for the metallic case, 
but does for the insulating case, in which finite energy is necessary 
to break a local singlet pair. 
\par

\begin{figure}
\epsfxsize=6cm
\epsfysize=2cm
\caption{
Charge density structure factor by $\Psi_{\rm Q}$ at half filling 
for various values of $U/t$. 
Inset shows $N(q)$ at $q=G$ (M point) as a function of coupling 
strength. 
Four system sizes are shown. 
$10^6$-$2\times10^6$ samples are used. 
}
\label{fig:udnqhf}
\end{figure}

\subsection{Results for less-than-half filling}
In the repulsive case $\tilde\Psi_{\rm Q}$ becomes inadequate as soon as 
the electron density leaves half filling, because the symmetry between 
d-sites and e-sites breaks. 
In the attractive case, however, $\Psi_{\rm Q}$ remains good even for 
less-than-half filling as will be shown, because $\uparrow$-site 
and $\downarrow$-site preserve the symmetry even for $n<1$. 
\par

First, we touch on the behavior of energy as a function of $\mu$ 
(not shown). 
For $n\gsim 0.5$ the feature of both $E_t/t$ and $E_U/U$ is qualitatively 
the same with at half filling (Insets of Fig.~\ref{fig:udetu}), whereas 
for densities as low as $n=0.24$ both $E_t/t$ and $E_U/U$ are always 
monotonically increasing functions of $\mu$. 
Thus, energy gain by $\Psi_{\rm Q}$ for small $n$ is ascribed only to 
the potential part even for small $|U|/t$. 
Anyway, energy is reduced from that of GWF for arbitrary values of 
$n$ and $U/t$ ($U<0$). 
\par

Shown in the insets of Figs.~\ref{fig:udeopt}(a) and (b) are the optimized 
values of $E_t/t$ and $E_U/U$ respectively, the behavior of which is 
basically the same with that at half filling. 
We have estimated $U_Q$ for all the available densities of $L=10$ 
in a similar manner to half filling, and showed it in Fig.~\ref{fig:BRT}. 
The value of $U_Q$ thus determined is weakly dependent on $n$: 
$|U_Q|/t$ increases subtly with decreasing $n$ from half filling, 
has a maximum about 9.2 around $n=0.5$, then slowly decreases and seems 
to tend to $|U|=W=8t$ in the low density limit. 
This behavior is qualitatively similar to that of GA discussed in 
Appendix A. 
Incidentally, it is known that in the low density limit an s-wave 
onsite pair is formed for any negative value of $U$ for the square 
lattice,\cite{Micnus} and is connected to s-wave superconducting states 
in finite densities. 
Nevertheless, accurate properties in this limit within the normal phase 
seem still unsettled. 
\par

\begin{figure}
\caption{
Critical values of metal-insulator transition in 2D attractive 
Hubbard model for the up- and down-spin binding wave function and for 
Brinkman-Rice transition in Gutzwiller approximation. 
The magnitude of circle indicates possible errors. 
For the former system-size dependence is negligible in this scale; 
in fact the results for $L=12$ ($n=0.5$ and $1.0$) are included among 
those for $L=10$. 
The crosses indicate the critical value of metal-insulator transition by 
DMFT with a half-ellipse DOS.\cite{Capone}
The boundary $U_{\rm m}$ of another instability arising in GA is also 
shown for 2D by a dash-dotted line. 
In the low-density limit, $U_{\rm m}/t\sim -4.9$. 
Since chemical potential is constant with respect to $n$ between the 
lines $U_{\rm m}$ and $U_{\rm BR}$, the phase there is unstable. 
Since DOS of the square lattice is flat at the band edges, the region of 
this instability does not spread to the vicinity of half filling, in 
contrast to the case 
of a half-ellipse DOS.\cite{Laloux} 
The explanation as to $U_{\rm m}$ and BRT will be given in Appendix A. 
}
\label{fig:BRT}
\end{figure}
\par 

Now, we go on to the correlation functions by $\Psi_{\rm Q}$. 
Shown in Fig.~\ref{fig:udcfpf}(a) is momentum distribution in both sides of 
$U_Q$ for some values of $n$. 
In every density definite discontinuities at $k_{\rm F}$ can be seen 
in the weak correlation side ($U/t=-8$ indicated by solid symbols), 
whereas in the strong coupling side ($U/t=-12$, open symbols) we can 
hardly find anomaly at $k_{\rm F}$. 
Thus, the features near the Fermi surface are basically accords with 
those at half filling. 
In Fig.~\ref{fig:udcfpf}(b) $N(q)$ and $P(q)$ is depicted for some values 
of $n$. 
Here, although the data in the weak correlation side ($|U|/t=8$) is 
employed, ones for $|U|>|U_Q|$ make no difference qualitatively. 
As for $N(q)$, as soon as electron density leaves half filling, $N(G)$ 
greatly diminishes and the maximum position moves to an incommensurate 
wave number. 
Accordingly, dominant antiferro CDW correlation is restricted to half 
filling. 
In contrast, the peak of $P(q)$ at $q=0$ remains sharp even for $n=0.24$; 
singlet-pairing correlation survives in the whole electron density. 
All these features are the same with those of GWF. 
Lastly concerning the gaps, the small-$|q|$ behavior of $N(q)$ and 
$S(q)$ is essentially independent of $n$, thus the same with that 
at half filling. 
Accordingly, $\Psi_Q$ is always gapless for charge degree of freedom, 
whereas a finite spin gap opens in the insulating phase. 
\par

\begin{figure}
\epsfxsize=6cm
\epsfysize=2cm
\caption{
(a) Momentum distribution function by $\Psi_{\rm Q}$ for four values of 
$n$ and $U/t=-8$ and $-12$. 
(b) Structure factors of charge density and onsite singlet pairing by 
$\Psi_{\rm Q}$ for four values of $n$ and $U/t=-8$. 
Inset shows spin structure factor on the ${\rm\Gamma}$-X 
line at half filling for the metallic (solid symbols) and insulating 
(open ones) regions, where the values of $U/t$ are $-9,-10,-11,-13,-16$ 
and $-32$ from above. 
In each panel $10^6$-$2\times10^6$ samples are used for each parameter 
set. 
}
\label{fig:udcfpf}
\end{figure}

\subsection{Discussion and comparison}
First of all, we compare the above results with those of 
DMFT.\cite{MetznerDMF,Capone} 
In DMFT there are two types of solutions---metallic ($|U|<|U_{\rm c2}|$) 
and insulating ($|U|>|U_{\rm c1}|$)--- in the normal phase, although 
details are different among different conditions or methods used 
to solve the Anderson impurity problem. 
In a intermediate regime $|U_{\rm c1}|<|U|<|U_{\rm c2}|$, where two 
solutions coexist and a kind of phase separation may arise, a 
metal-insulator transition takes place.
Indicated in Fig.~\ref{fig:BRT} by crosses is that critical value, 
which seems to decreases monotonically with decreasing $n$. 
The fact that the values of VMC and DMFT broadly coincide corroborates 
the existence of a metal-insulator transition, which brings about a change 
in the mechanism of superconductivity, even in a realistic dimension. 
\par 

The above unstable phenomenon in the intermediate correlation strength 
had been reported for the half-filled RHM in a magnetic field by 
DMFT.\cite{Laloux} 
According to it, in the intermediate regime of $U$ there appear 
a jump and a hysterises in the magnetization curve. 
This instability was considered as an analogue of the one arising in 
GA (equivalently GWF in this case), which we study in Appendix A. 
However, there is an evident difference in that the magnetization in DMFT 
does not saturate after this discontinuity. 
Here, we point out that this instability found in DMFT has nothing 
to do with the one in GWF. 
To begin with, recall that the range of instability in GWF necessarily 
includes $n=0$, namely total energy is a linear function of $n$ 
in a finite range near $n=0$, as shown in Fig.~\ref{fig:GEvsn}. 
On the other hand, in Fig.~\ref{fig:udEvsn} we show the same quantity 
by $\Psi_Q$. 
In this case the total energy is a concave function of $n$ in the 
low density regime for any coupling strength, namely 
$1/\chi_c=\partial^2E/\partial n^2>0$. 
Thus, there is no instability for $n\sim 0$. 
It follows that the instability brought by GWF is spurious at least 
for small $n$. 
Incidentally, $E/t$ for $L=10$ shows a tendency to have a cusp-like 
winding at $n=0.6$ (also slightly at $0.32$) and be convex around 
$n=0.7$ for $L=10$, similarly for $L=8$ a cusp at $n=0.5625$ and 
convexity at $n\sim 0.7$. 
This is probably caused by the finite system sizes,\cite{noteinstab} 
and has nothing to do with the instability expected in the critical 
regime by DMFT.\cite{notePS} 
\par

\begin{figure}
\epsfxsize=6cm
\epsfysize=2cm
\caption{
Energy of $\Psi_Q$ measured from the value at $U=-\infty$ 
($E_\infty=Un/2$) as a function of $n$ for several values of $U/t$. 
Symbols denotes the VMC data, some of which are fit linearly in 
low density as shown by dashed lines. 
To check the system-size dependence in a high-density regime, 
the data for $L=8$ are simultaneously plotted with those for 
$L=10$. 
}
\label{fig:udEvsn}
\end{figure}

Now, returning to Fig.~\ref{fig:renorm}, we compare $Z$ of $\Psi_Q$ 
with that calculated from the self energy in DMFT. 
The behaviors of $Z$'s roughly coincide at $U\rightarrow 0$ and 
$U\sim U_Q$, but does not for the intermediate values of $U$. 
It is probable that some intrinsic effect of electron correlation 
is not sufficiently introduced in $\Psi_Q$ for the metallic regime. 
Indeed, $Z$ of $\Psi_Q$ is very close to that of GWF for small $|U|$.
\par

Then, we look at the improvement in $E/t$ on GWF. 
Figure \ref{fig:udDE} shows the energy decrement by $\Psi_{\rm Q}$ 
for some $L$'s at half filling and simultaneously for some $n$'s 
for $L=10$. 
In every case, $\Delta E$ increases exponentially at first with 
increasing $|U|/t$, reaches a maximum around $U/t=-12$, then decreases 
as $t/|U|$. 
The improvement in the metallic regime is certainly small. 
The chief reason of the maximum at $U/t\sim -12$ is the abrupt change 
of $E-Un/2$ in GWF, as seen in Fig.~\ref{fig:totEGWF}. 
As for the transition of $\Psi_{\rm Q}$, the change of $E-Un/2$ 
around $U_Q$ is unexpectedly gentle and does not cause a conspicuous 
effect on $\Delta E$. 
Moreover, the large system-size dependence is attributed mostly to GWF. 
Accordingly, one has to be deliberate in adopting $E(\Psi_{\rm G})$ 
as a standard of the normal-state energy, for example, when one 
estimates a condensation energy. 
\par 

\begin{figure}
\epsfxsize=6cm
\epsfysize=2cm
\caption{
Energy difference between GWF and $\Psi_Q$ as a function of 
$|U|/t$. 
At half filling data for four system sizes are plotted with solid 
symbols. 
The value of $|E|/t$ itself decreases with increasing $L$ for both 
$\Psi_{\rm G}$ and $\Psi_Q$, but the decrement of $E(\Psi_{\rm G})$ 
is thrice larger than that of $E(\Psi_Q)$ in comparing $L=12$ with 
$L=6$. 
For $U/t=-12$, which gives the maximum, $\Delta E$ in the thermodynamic 
limit is extrapolated as 0.204 by second-order polynomial fit from the 
data up to $L=20$ for GWF and up to $L=12$ for $\Psi_Q$. 
Simultaneously, the data of three densities for $n<1$ ($L=10$) are 
shown with open symbols. 
The dashed [dash-dotted] line is an arbitrary one proportional to 
$t/|U|$ [$\exp(-t/U)$]. 
}
\label{fig:udDE}
\end{figure}

\section{Summary}

In this paper we have studied normal-state properties of the attractive 
Hubbard model, especially in 2D, based on a series of variational Monte 
Carlo calculations. 
Main results are itemized as follows. 
\par

(1) In the Gutzwiller wave function a precipitous quantitative 
change in various properties takes place around an interaction strength 
$|U|=|U_{\rm co}|$ somewhat larger than the band width. 
In the region of $|U|>|U_{\rm co}|$ the onsite pairing correlation 
grows, and the effective mass of electrons becomes large. 
In both sides of $U_{\rm co}$, however, a clear Fermi surface exists 
and there is no excitation gap. 
Hence, this change in GWF is a metal-to-metal crossover, and is 
distinguished from the Brinkman-Rice metal-insulator transition, 
although the value of $U_{\rm co}$ is similar to $U_{\rm BR}$. 
Furthermore, we have confirmed with better statistics that BRT is 
an artifact of the Gutzwiller approximation, and does not arise 
in finite lattice dimensions. 
On the other hand, we have mentioned GWF itself has unphysical 
features to be refined in correlation functions and the phase 
stability. 
\par

(2) An improved wave function $\Psi_Q$ with a binding between 
$\uparrow$- and $\downarrow$-spins in adjacent sites has been introduced 
by considering the perturbation expansion in the strong correlation 
limit. 
This wave function succeeds to the merit of GWF but improves its weak 
points; $\Psi_Q$ undergoes a substantial metal-insulator transition 
at $|U_Q|$ roughly of the band width, which is favorably compared with 
the result of dynamical-mean-field approximations. 
We have constructed a phase diagram in the $U/t$-$n$ plane; in the 
weak-coupling regime $|U|<|U_Q|$ Fermi surface is clearly recognized 
and there is no excitation gap, whereas for $|U|>|U_Q|$ the jump of 
$n(k)$ at $k=k_{\rm F}$ vanishes and there appears gap behavior 
in spin degree of freedom in accordance with QMC.\cite{Randeria} 
The state in the insulating regime is different from those of DMFT 
and of GA in that kinetic energy does not vanish. 

(3) We have considered a Jastrow-type long-range correlation factor 
as an alternative improvement. 
At half filling this wave function undergoes a first-order transition 
to the antiferro CDW at a finite value of $U$, whereas for incommensurate 
fillings, such a transition does not occur. 
In the metallic regime weak attractive intersite correlation as well as 
appropriate attractive onsite correlation slightly reduces the energy, 
but the state is essentially the same with GWF. 
Once a CDW state is realized, singlet pairing correlation is severely 
suppressed in this wave function. 
\par

In the remainder, we make a couple of short discussions with future 
studies in mind. 
As for the metal-insulator transition the result of $\Psi_Q$ is 
consistent by and large with that of DMFT. 
Meanwhile, a recent study by a non-self-consistent $T$-matrix 
approximation\cite{MetznerTM} concluded that in case of 2D a 
pseudogap in the single-particle spectral weight is induced even 
in a weak-interaction regime $|U|\ll W$ due to the pair 
fluctuation,\cite{Vilk} if the temperature is sufficiently close 
to $T_{\rm c}$. 
In fact, a sign of pseudogap is also recognized for $|U|<W/2$ in QMC 
data,\cite{TR,Singer} although this feature does not seem restricted 
to the 2D square lattice in seeing similar QMC results obtained for 2D 
triangular and 3D simple cubic lattices.\cite{Santos} 
To study this issue $\Psi_Q$ has room for improvement, because pairing 
correlation as well as pairing fluctuation is not introduced 
deliberately for $|U|\ll|U_Q|$ as mentioned in \S4.3. 
\par

Next, we refer to the difference between AHM and RHM. 
A common aspect among the three normal-state wave functions studied 
is that many properties in AHM only weakly depend on electron density 
except for the CDW correlation. 
This aspect is also common to a superconducting state in AHM,\cite{Y} 
but in sharp contrast with that of RHM, in which Mott transition is 
limited to the high-density regime especially at half filling. 
In addition, we have shown that $\Psi_{\rm J}$ hardly improve $\Psi_{\rm G}$ 
for AHM unlike for RHM, and that for $n<1$ the behavior of correlation 
functions by GWF is distinct between AHM and RHM. 
Moreover, in the insulating regime $\Psi_Q$ is gapless in charge 
degree of freedom, but is gapped in spin one for any electron density. 
This corresponds to the accepted conception that RHM at half filling 
has a Hubbard gap in charge part but is gapless in spin part. 
Nevertheless, if a similar insulating state is assumed for $n<1$ 
in RHM, the gap behavior on spin and charge must be reversed from 
half filling. 
For this reason, it is not appropriate to discuss RHM in $n<1$ on 
the analogy of AHM with $h=0$; one should directly study RHM. 
Since $\Psi_Q$ is not properly extended to less-than-half filling for 
RHM, it is important to study a wave function in which intersite 
correlations are systematically introduced.\cite{Otsuka} 
\par

In this paper we have not touched on superconducting states, although 
in the real ground state of AHM an s-wave superconducting order prevails 
all over the parameter plane, and an antiferro CDW order coexists 
particularly at half filling. 
Actually, trial wave functions with such orders further reduce 
the energy; we will treat this subject in a coming publication.\cite{Y} 

\section*{Acknowledgements}
The author thanks Yoshiaki \=Ono and Dai S.~Hirashima for useful 
discussions. 
This work is partly supported by a Grant-in-Aid of Scientific Research 
on Priority Areas (B) ``New Properties of Matter due to Ordering and 
Fluctuation of Electron Orbitals" from Ministry of Education, Science, 
Sports and Culture. 
A part of computation was done with the facilities of Supercomputer 
Center, Institute for Solid State Physics, University of Tokyo. 
\par

\appendix
\section{On Brinkman-Rice Transition}
In a previous paper\cite{YS1} we discussed the absence of the 
Brinkman-Rice metal-insulator transition (BRT) for RHM with the VMC 
results. 
Afterward, the exact formula for GWF in $D=1$ and $D=\infty$ ($D$: lattice 
dimension) were diagrammatically derived;\cite{MV,dinf} for $D=1$ the 
absence of BRT is proven, whereas for $D=\infty$ the exact formula 
coincide with the GA formula, so that BRT exists. 
As for $2\le D<\infty$, there is no proof of the absence, but van 
Dongen \etal\cite{dinf} concluded the absence of BRT by making a 
couple of plausible assumptions about the scaling properties of 
spin-spin correlation function. 
In this Appendix we reconsider it with refined VMC data for $U<0$, 
and refer to the dependence on lattice dimension. 
A part of the discussions on this transition was already mentioned 
in Ref.~\citen{Thesis}.
\par 

First of all, let us look at the Gutzwiller approximation (GA) 
for $U<0$\cite{GAA} and $H=0$ briefly, which are obtained in a similar 
fashion to the repulsive case.\cite{Gutz,Voll} 
According to GA, total energy is given by $E=q\varepsilon_0+Ud$ 
for $n_\uparrow=n_\downarrow=n/2$, where $\varepsilon_0$ is the 
noninteracting energy per site for given $n$, and $q$ is the 
renormalization factor in GA: 
\begin{equation}
q=\frac{\left[\sqrt{\left(\frac{n}{2}-d\right)(1-n+d)}
              +\sqrt{\left(\frac{n}{2}-d\right)d}\right]^2}
       {\frac{n}{2}\left(1-\frac{n}{2}\right)}.
\end{equation} 
$E$ has to be minimized with respect to $d$. 
With increasing $|U|/t$ electrons more and more form onsite pairs, 
and singly occupied sites finally vanishes at a critical value 
$U=U_{\rm BR}$, where $d=n/2$, $q=0$ and $\gamma=1/g=0$. 
Thus, electrons are completely localized and kinetic energy vanishes. 
In contrast to the repulsive case, BRT takes place at arbitrary 
electron density for $U<0$. 
In Fig.~\ref{fig:BRT} the critical value $|U_{\rm BR}|$ for the square 
lattice is plotted together.
At half filling the critical value has the same magnitude with the 
repulsive case: $U_{\rm BR}/t=-8|\varepsilon_0|/t$ ($=-12.969\cdots$); 
with decreasing $n$, $|U_{\rm BR}|$/t once increases slightly, has a 
maximum at $n\sim 0.62$, then decreases. 
The behavior for $n\rightarrow 0$ is singular. 
\par 

Before examining BRT, we touch on another instability arising in GWF. 
In connection with magnetic properties of $^3$He and heavy-fermion 
systems, some papers treated RHM in a magnetic field $H$. 
Magnetization curve $m(H)$ obtained by GA\cite{Voll} has a discontinuous 
jump to the full moment at a value of $H$ for 
$U_{\rm m}<U<U_{\rm BR}$ with $U_{\rm m}=\xi U_{\rm BR}$ 
($\xi=0.38$ for 2D square lattice). 
Namely, there exists an unstable region of $m$ near $m=1$ for intermediate 
values of $U/t$. 
This fact is interpreted for AHM through the canonical transformation 
eq.~(\ref{eq:canon}) as follows: For $|U_{\rm m}|<|U|<|U_{\rm BR}|$ 
there exists a finite region for small $n$ where chemical potential 
$\zeta$ is constant, namely particle number is variable. 
We show the boundary of this unstable region also in Fig.~\ref{fig:BRT}. 
Unlike BRT, this instability does not stem from GA but from GWF, because 
quantitatively similar jumps are observed in VMC results for 
$\Psi_{\rm G}$.\cite{YT} 
This instability is confirmed for AHM as the constant chemical potential 
[$\zeta=(\partial E/\partial n)_L$] as shown in Fig.\ref{fig:GEvsn}, 
where total energy of GWF by VMC calculations are plotted versus $n$. 
Total energy is a linear function of $n$ for small $n$, and deviates 
roughly at GA's critical values indicated by arrows. 
As long as this instability is concerned, GA is broadly correct. 
Anyway, this instability near $n=0$ is an artifact of GWF and removed 
in $\Psi_Q$ as discussed in \S4.
\par

\begin{figure}
\caption{
Energy of GWF measured from the value at $U=-\infty$ ($E_\infty=Un/2$) 
as a function of $n$ for several values of $U/t$. 
Symbols denote the VMC data, which are fit linearly in low density 
as shown by dashed lines. 
Arrows indicate GA's critical values of $n$ (at $U_{\rm m}$) under 
which chemical potential is constant. 
}
\label{fig:GEvsn}
\end{figure}

Now, we return to BRT. 
Since this transition is caused by the behavior of kinetic and 
potential energies for small-$\gamma$ region, we consider the 
expansion in $\gamma$, first. 
In GA the leading terms of both $E_t/t$ and $E_U/U$ for 
$\gamma\rightarrow 0$ are linear in $\gamma$ for arbitrary $n$; 
at half filling, in particular, they are simply written as, 
$$
\frac{E_t}{t}=4\frac{\varepsilon_0}{t}\frac{\gamma}{(1+\gamma)^2}
=4\frac{\varepsilon_0}{t}\gamma(1-2\gamma+\cdots),\quad
\frac{E_U}{U}=\frac{1}{2}\frac{1}{1+\gamma}=\frac{1}{2}(1-\gamma+\cdots).
$$
The coefficient of the leading term of total energy $E_{\rm tot}=E_t+E_U$ 
changes its sign at $|U|=8|\varepsilon_0|t$. 
Thereby, it is determined whether the minimum is situated at $\gamma=0$ 
(namely localized) or not. 
For $n\ne 1$, the situation does not change. 

Now, let us consider GWF without approximations. 
For $\gamma(=1/g)\rightarrow 0$, GWF is expanded as, 
\begin{eqnarray}
\Psi_{\rm G}&=&\prod_j\left[1-(1-g)d_j\right]\Phi_{\rm F} 
             =g^N\prod_j\left[\gamma\bar d_j+d_j\right]\Phi_{\rm F}
\\
            &=&g^N\gamma^{N-N_{\rm e}/2} 
\left[           \Psi_{\rm G}^{(0)} 
                +\gamma\Psi_{\rm G}^{(1)} 
                +\gamma^2\Psi_{\rm G}^{(2)}+\cdots 
                +\gamma^{N_{\rm e}/2}\Psi_{\rm G}^{(N_{\rm e}/2)} 
\right],\qquad\quad 
\end{eqnarray}
with $d_j=n_{j\uparrow}n_{j\downarrow}$ and 
$\bar{d_j}=1-n_{j\uparrow}n_{j\downarrow}$.
Here, $\Psi_{\rm G}^{(\ell)}$ is the state with 2$\ell$ singly-occupied 
sites:
\begin{equation}
\Psi_{\rm G}^{(\ell)}=\sum_{\{d:N_{\rm e}/2-\ell\}}\left[
\prod_j^{(N-N_{\rm e}/2+\ell)}\bar d_j
\prod_{k(\ne j)}^{(N_{\rm e}/2-\ell)}d_k\right]\Phi_{\rm F},
\end{equation}
where all the configurations with $N_{\rm e}/2-\ell$ d-sites 
are summed up. 
If we assume the expectation values of kinetic and potential energies 
by $\Psi_{\rm G}$ can be expanded with respected to $\gamma$, 
they are formally written as
\begin{equation}
E_t=\frac{\langle\Psi_{\rm G}|{\cal H}_t|\Psi_{\rm G}\rangle}
{\langle\Psi_{\rm G}|\Psi_{\rm G}\rangle}
=t\left[\gamma K_1+\gamma^2 K_2+O(\gamma^3)\right]
\label{eq:expEt}
\end{equation}
\begin{equation}
E_U=\frac{\langle\Psi_{\rm G}|{\cal H}_U|\Psi_{\rm G}\rangle}
{\langle\Psi_{\rm G}|\Psi_{\rm G}\rangle}
=U\left[P_0+\gamma^2P_2+O(\gamma^4)\right].
\label{eq:expEU}
\end{equation}
Here, $K_1=T_1/N_0$, $K_2=T_2/N_0$, $P_0=U_0/N_0$ and 
$P_2=(U_2-U_0N_1/N_0)/N_0$,
with
$$
N_\ell=\langle\Psi_{\rm G}^{(\ell/2)}|\Psi_{\rm G}^{(\ell/2)}\rangle,
\quad
U_\ell=\langle\Psi_{\rm G}^{(\ell/2)}|{\cal H}_U|
\Psi_{\rm G}^{(\ell/2)}\rangle/U,
$$
$$
T_1=\left[\langle\Psi_{\rm G}^{(0)}|{\cal H}_t|\Psi_{\rm G}^{(1)}\rangle
    +\langle\Psi_{\rm G}^{(1)}|{\cal H}_t|\Psi_{\rm G}^{(0)}\rangle\right]/t,
\quad
T_2=\langle\Psi_{\rm G}^{(1)}|{\cal H}_t|\Psi_{\rm G}^{(1)}\rangle/t.
$$ 
Henceforth, we concentrate on the half-filled band, because system-size 
dependence can be checked and the situation does not change for $n<1$, 
in addition to another merit that the discussion here is directly 
connected to the conventional repulsive case with $n=1$ and $h=0$ through 
eq.~(\ref{eq:symGWF})
\par

For finite systems the situation is simple, because the expansion 
coefficients $K_\ell$'s and $P_\ell$'s are finite, namely the 
expansion by $\gamma$ eqs.~(\ref{eq:expEt}) and (\ref{eq:expEU}) 
are pertinent. 
This is seen in Figs.~\ref{fig:EGINF}(a) and (b), where $E_t/t$ and 
$E_U/U$ are plotted versus $\gamma (\sim 0)$ for some values of $L$. 
Since the leading terms of $E_t/t$ and $E_U/U$ for $\gamma\rightarrow 0$ 
are linear and quadratic, respectively, $E_{\rm tot}$ has a minimum 
at $\gamma=-tK_1/(UP_2)$ for $|U|\gg t$. 
Thus, $\gamma$ does not vanish as long as $|U|/t$ is finite: BRT is absent. 
\par

\begin{figure}
\epsfxsize=6cm
\epsfysize=2cm
\caption{
(a) Kinetic and (b) potential energy of GWF as a 
function of $\gamma$ near $\gamma=0$. 
System-size dependence is compared. 
The results of GA are also shown by a dash-dotted line. 
In (a) the leading linear term for $L=\infty$ is shown by a shadowed
line. 
Its slope $K_1$ (See Table I) is probably a half of the value by GA: 
$4\varepsilon_0/t$. 
The dashed line smoothly connects the data for $L=20$. 
Error bars are attached only for $L=20$ and 24 for clarity. 
In (b) the data for $L=6$-20 are fitted by quadratic curves (solid line). 
As $L$ increases, the range of $\gamma$ where $E_U/U$ obeys 
$n/2-{\rm const.}\times\gamma^2$ shrinks; dashed lines connecting 
data points for $L=24$ are guide for eyes. 
$5\times10^4$-$10^7$ samples are used. 
}
\label{fig:EGINF}
\end{figure}

Nevertheless, the situation is not so simple for the thermodynamic limit. 
Note that in Fig.~\ref{fig:EGINF} the behavior of $E_t/t$ for $\gamma\sim 0$ 
seems almost independent of $L$, whereas that of $E_U/U$ is highly dependent. 
Therefore, let us consider the coefficients $K_1$, $K_2$ and $P_2$, 
which are also obtainable by VMC calculations. 
In Fig.~\ref{fig:coeff}(a) the coefficients thus obtained, which are 
naturally independent of $\gamma$, are shown for $L=10$ and for some 
values of $\gamma$. 
The values thereby and similarly for different $L$'s are 
plotted as a function of $1/L$ in Fig.~\ref{fig:coeff}(b). 
$K_1$ only weakly depends on $L$ and is well fitted linearly, whereas 
$K_2$ and $P_2$ largely depend on $L$ and diverge as $-a-b/(1/L)^c$ 
for $\gamma\rightarrow 0$.
It follows the expansion form eqs.~(\ref{eq:expEt}) and (\ref{eq:expEU}) 
is inadequate. 
This situation by VMC analysis is quite common to the 1D case.\cite{Thesis} 
On the other hand, according to the exact analytic results 
for the 1D repulsive model,\cite{MV} the leading term for $E_U/U$ 
has logarithmic correction as $g^2\ln(1/g)$, while 
$E_t/t=2g\varepsilon_0/t$ for $n=1$ and $g\rightarrow 0$. 
Thus, it is probable that such a nonanalytic factor appears also 
in higher dimensions. 
Of course, logarithmic correction does not affect the absence of BRT,
though. 
\par

\begin{figure}
\epsfxsize=6cm
\epsfysize=2cm
\caption{
(a) Low-order expansion coefficients of 2D GWF for 
$\gamma\rightarrow 0$ and $n=1.0$ calculated by a VMC method. 
Incidentally, $P_0=n/2$.
The least square fit yields $K_1=-3.24\pm 0.01$, $K_2=-30.90\pm 0.09$, 
and $P_2=-5.83\pm 0.02$. 
Accuracy of calculations becomes high around the maximum of 
$|\Psi_{\rm G}^{(1)}|^2$, \eg $\gamma\sim 0.045$ for $L=10$ 
as seen in the inset of (b). 
(b) System-size dependence of low-order expansion coefficients. 
$K_1$ for $L=\infty$ is estimated as $-3.235$ by the second-order 
polynomial fit. 
The other curves are well fitted by the formulae $-a-b/(1/L)^c$, 
where $a=3.79, b=0.932, c=1.46,$ for 
$K_2$ and $a=0.86, b=0.173, c=1.46$ for $P_2$. 
The power $c$ seems common to $K_2$ and $P_2$ irrespective of 
lattice dimension (see Table I). 
Inset shows the weight of $\Psi_{\rm G}^{(\ell)}$, that is 
$
\gamma^{2(\ell-N_{\rm e})}
\langle\Psi_{\rm G}^{(\ell)}|\Psi_{\rm G}^{(\ell)}\rangle/
\langle\Psi_{\rm G}|\Psi_{\rm G}\rangle, 
$
as a function of $\gamma$. 
}
\label{fig:coeff}
\end{figure}

Instead of the expansion, let us consider $E_U/U$ directly, 
which is not a divergent quantity. 
Assume the leading power of $E_U/U$ is continuous: 
$E_U/U$=$E_\infty/U-{\rm const.}\times\gamma^\alpha$. 
Then, the optimized value of $\gamma$ is nonzero for finite $U/t$ 
if $\alpha>1$, since the leading term of $E_t/t$ is linear in 
$\gamma$ due to the above discussion. 
In Figs.~\ref{fig:Edir2D}(a) and (b), $E_t/t$ and $E_U/U$ are 
depicted for several small values of $\gamma$ as a function of $1/L^2$. 
Since the values changes rapidly with increasing $L$ and we 
do not know a priori the system-size dependence, reliable 
extrapolation is not anticipated from the raw data.\cite{note1D} 
Therefore, we consider the ratio between the values of different 
$\gamma$; here we use $\gamma=1/201$ as a reference. 
In Figs.~\ref{fig:Erat2D}(a) and (b) such ratios are plotted for 
$E_t/t$ and $(E_\infty-E_U)/U$, respectively. 
In this case $E_\infty/U=0.5$.
In this plot the system-size dependence is smaller, so that the 
behavior in the thermodynamic limit can be estimated to some extent. 
The crosses on the vertical axis indicate the values if $E_t/t$ 
or $(E_\infty-E_U)/U$ is precisely proportional to $\gamma$, 
namely $1/(201\gamma)$. 
Similarly, the dots indicate the case of $\gamma^2$: $1/(201\gamma)^2$. 
In Fig.~\ref{fig:Erat2D}(a) all the curves substantially point to 
the crosses; $E_t/t$ is proportional to $\gamma$, as expected. 
On the other hand, in Fig.~\ref{fig:Erat2D}(b) the extrapolated values 
seems by far nearer to the dots than to the crosses especially for 
small values of $\gamma$; the effective power $\alpha\sim 2$ (at least $>1$). 
Some deviation from purely linear or quadratic behavior 
for large $L$ may be attributed to possible nonanalytic factors, 
besides using small but finite magnitude of $\gamma$. 
Thus, we can confidently conclude BRT is absent not only in 1D 
but in 2D. 
\par

\begin{figure}
\epsfxsize=6cm
\epsfysize=2cm
\caption{
System-size dependence of (a) $E_t/t$ and (b) $E_U/U$ for several small 
values of $\gamma$ on the square lattice. 
Symbols are common in (a) and (b). 
$5\times10^5$($L=24$)-$10^7$($L=6$) samples are averaged. 
}
\label{fig:Edir2D}
\end{figure}

\begin{figure}
\epsfxsize=6cm
\epsfysize=2cm
\caption{
System-size dependence of (a) $E_t/t$ and (b) $[E_U(\gamma=0)-E_U]/U$ 
for several small values of $\gamma$ normalized by the value at 
$\gamma=1/201$.
Crosses (dots) on the vertical axis indicate the values assuming 
$E_t(\gamma)$/t or $[E_\infty-E_U(\gamma)]/U$ is precisely proportional 
to $\gamma$ ($\gamma^2$).
}
\label{fig:Erat2D}
\end{figure}
\par

Similar analyses have been carried out for the simple cubic lattice. 
Compared with 2D in Fig.~\ref{fig:Edir2D}, the behavior of $E_t/t$ 
and $E_U/U$ in 3D (not shown) is qualitatively similar, but the rapid 
change for $\gamma\rightarrow 0$ occurs at a smaller value of $L$ 
(or a large value of $\gamma$) in both energy components. 
In Fig.~\ref{fig:Erat3D} the ratios corresponding to Fig.~\ref{fig:Erat2D} 
are shown for 3D. 
Although the feature is substantially the same with that in 2D, 
smaller $\gamma$ is needed to single out the behavior of leading terms. 
To confirm this aspect quantitatively, we compare the expansion 
coefficients among 1D, 2D and 3D lattices in Table I. 
As for the kinetic part, $K_1$ scarcely changes with the system size 
in each dimension, and the relation $K_1=2\varepsilon_0/t$ holds 
not only for 1D, but probably for other finite dimensions
($\varepsilon_0/t=-4/\pi, -16/\pi^2, -2.0048\cdots$ for 
$D=1,2$ and 3, respectively). 
This point is in sharp contrast to the $1/D$ expansion discussed below.
On the other hand, with increasing lattice dimension not only the 
absolute values of $P_2$ and $K_2$ increase, but their divergent 
behavior for $L\rightarrow\infty$ becomes more remarkable. 
For instance, compare the power $c$ in Table I.
Consequently, each energy component by GWF quantitatively approaches 
the value of GA with increasing $D$, but the qualitative features for 
$\gamma\ (g)\rightarrow 0$ remain distinct.
\par

\begin{figure}
\epsfxsize=6cm
\epsfysize=2cm
\caption{
System-size dependence of (a) $E_t/t$ and 
(b) $[E_U(\gamma=0)-E_U]/U$ normalized by each value at 
$\gamma=1/201$ for several small values of $\gamma$.
Crosses and dots on the vertical axis indicate the same with 
Fig.~\ref{fig:Erat2D}. 
Samples as many as $2.5\times10^5$($L=10$)-$5\times 10^6$($L=4$) are 
averaged. 
}
\label{fig:Erat3D}
\end{figure}
\par

\begin{table}
\caption{
Expansion coefficients of $E_t/t$ and $E_U/U$ for 1D, 2D and 3D 
lattices.
In the last line but one the values of $2\varepsilon_0/t$ are entered 
for comparison with $K_1$. 
Entered in the last line are the common powers $c$ for $P_2$ and $K_2$ 
in the fitting form of $-a-bL^c$. 
For 1D, the data in ref.~\cite{Thesis} for RHM are adopted; the values 
are adjusted to the attractive case through eq.~(3.5): 
$K_\ell=\tilde K_\ell$ and $P_2=-\tilde P_2$ from the original data 
($\tilde K_\ell, \tilde P_2$). 
They are estimated from fewer samples ($5\times 10^4$-$10^5$) with 
periodic boundary condition. 
}
\label{table:coeff}
\begin{center}
\begin{tabular}{@{\hspace{\tabcolsep}\extracolsep{\fill}}rrrr|rrrr|rrrr}
\hline\hline
\multicolumn{4}{c|}{1D chain}  &
\multicolumn{4}{c|}{2D square} &
\multicolumn{4}{c}{3D simple cubic} \\ \hline
$L$&$K_1$&$K_2$&$P_2$&$L$&$K_1$& $K_2$&$P_2$&$L$& $K_1$& $K_2$& $P_2$ \\
\hline                                                              
 30&-2.57&-5.76&-1.87&  6&-3.32& -16.2& -3.1& 4 & -4.06 &  -33&  -4.7  \\
 50&-2.53&-6.71&-2.11&  8&-3.28& -23.3& -4.4& 6 & -4.05 &  -88& -12.2  \\
 70&-2.55&-7.30&-2.27& 10&-3.27& -30.9& -5.8& 8 & -4.01 & -186& -25.8  \\
 90&-2.55&-7.71&-2.37& 12&-3.27& -39.1& -7.3&10 & -4.01 & -338& -46.8  \\
130&-2.53&-8.67&-2.63& 14&-3.25& -47.7& -8.9 \\
170&-2.58&-9.05&-2.74& 16&-3.25& -57.0&-10.6 \\
210&-2.59&-9.08&-2.76& 18&-3.24& -67.1&-12.5 \\
   &     &     &     & 20&-3.25& -78.7&-14.6 \\ 
   &     &     &     & 24&-3.25&-102.8&-19.0 \\ \hline\hline
$2\varepsilon_0/t$&-2.546&&&$2\varepsilon_0/t$&-3.242&&&$2\varepsilon_0/t$ & -4.010 & & \\ \hline
$c$&&\multicolumn{2}{c|}{0.12}&$c$&&\multicolumn{2}{c|}{1.54}&$c$&&\multicolumn{2}{c}{2.76} \\ \hline
\end{tabular}
\end{center}
\end{table}

This point has been already argued by the $1/D$ expansion analysis.\cite{dinf}
In this approach the starting point is $D=\infty$, where 
GA becomes the exact estimation of GWF, and properties for finite 
dimensions are improved by $1/D$ (and $1/D^2$ etc.) corrections on GA. 
Owing to this correction, most properties are quantitatively 
improved even in 1D. 
However, the exception is the behavior for $\gamma\rightarrow 0$. 
In $1/D$ expansion, the leading terms of the energy components for 
$g\rightarrow0$ are calculated as $E_t/t=c_1g\varepsilon_0/t$ and 
$E_U/U=c_2g$, where 
$$
c_1=4\left[1-\frac{1}{2D}\left(\frac{\varepsilon_0}{t}\right)^2
-O(D^{-2})\right],\quad
c_2=\frac{1}{2}\left[1-\frac{1}{D}\left(\frac{\varepsilon_0}{t}\right)^4
-O(D^{-2})\right].
$$
This indicates that $1/D$ corrections only reduce the coefficient 
and $E_U/U$ remains linear in $g$ for small $g$. 
Therefore, this approach does not get rid of BRT, but only increases 
the value of $|U_{\rm BR}|$. 
In this connection it was pointed out that $D=\infty$ is singular 
as to the properties for $g\rightarrow 0$, so that $1/D$ expansion 
is inappropriate as long as near BRT.\cite{dinf} 
Similarly, the coefficient $c_1$ is reduced by $1/D$ corrections, 
which means different values of $c_1$ for different $D$'s. 
However, the value of $c_1$ by VMC is constant, that is 2, for the 
three realistic dimensions. 
Thus, $c_1$ also seems singular at $D=\infty$. 
\par

Summarizing, BRT is restricted to the infinite dimension, and never 
takes place in finite dimensions. 
$D=\infty$ is singular for the properties in $\gamma\ (g)\rightarrow 0$; 
the linear coefficients in the $\gamma$-expansion of both $E_t/t$ and 
$E_U/U$ have discontinuity at $D=\infty$, specifically as 
$2\varepsilon_0/t\rightarrow 4\varepsilon_0/t$ and $0\rightarrow 1/2$, 
respectively. 
It is a remaining issue how logarithmic corrections, which are not 
relevant in 1D, affect the power counting, if any for $2\le D<\infty$. 
\par

\section{Possibility of long-range orders in GWF} 

In this Appendix we study the singular behavior of correlation 
functions calculated with GWF as a supplement to \S3.3. 
To check the system-size dependence we have plotted the inverse of 
$N(G)$, $P(0)$ and $S(G)$ versus $1/L^2$ ($L=6$-20) for several values 
of $\gamma$ (not shown), and found that they are smoothly extrapolated 
toward $L=\infty$ by polynomials. 
Figure \ref{fig:CFinf} shows the values thereof as well as the original 
finite-size data versus $\gamma$. 
From this figure we can read a couple of facts. 
First, both $N(G)$ and $S(G)$ have very little system-size dependence 
for large values of $\gamma$, while they come to have appreciable 
size dependence with decreasing $\gamma$. 
The value of $\gamma$ at which system-size dependence becomes remarkable 
approximately coincides with $U=U_{\rm co}$ ($\gamma\sim 1.2$); 
simultaneously the system-size dependence changes its tendency from 
convergent to divergent (not shown). 
Thus, the crossover of the states also affects the correlation functions. 
Second, although $N(G)$ or equivalently $P(0)$ does not diverge for 
finite coupling strength, in the $\gamma\rightarrow 0$ limit the 
extrapolated inverse quantity of $L\rightarrow\infty$ is likely to vanish. 
Namely, $N(G)$ and $P(0)$ diverge in the limit of $\gamma=0$ or 
$|U|/t=\infty$. 
On the contrary, the order parameter $O_{\rm CDW}$ defined in \S2.1
measured in the same VMC sweep is always zero within statistical errors 
for any values of $\gamma$ and $L$. 
From these results we reach a conclusion that GWF does not possess 
CDW and singlet superconducting long-range order even in this limit. 
This point resembles the feature of 1D Heisenberg model. 
Finally, concerning $S(q)$, it is natural that the structure in $S(q)$ 
vanishes and its amplitude fades out with increasing $|U|/t$, according 
to the diminution of isolated spins. 
\par

\begin{figure}
\epsfxsize=6cm
\epsfysize=2cm
\caption{
Dependence of (a) $1/N(G)$ [or equivalently 1/2$P(0)$] and (b) $S(G)$ 
on a variational parameter $\gamma$ at half filling for various system 
sizes and extrapolated values for $L=\infty$, which are obtained by 
polynomial fit. 
Insets are magnification for small-$\gamma$ region. 
Solid lines are fitted curves by proper polynomials. 
Dashed lines for $L=\infty$ are guide for eyes. 
$2.5\times10^5(L=20)$-$10^6(L=6)$ samples are used for each parameter 
set. 
}
\label{fig:CFinf}
\end{figure}

To see the system-size dependence in less-than-half filling, we take up 
a quarter filling. 
As seen in Fig.~\ref{fig:cfqf}(a), $N(q)$ is enhanced for any $q$ 
by attractive $\gamma$ also in this density. 
On the path ${\rm \Gamma}\rightarrow{\rm X}\rightarrow{\rm M}
\rightarrow{\rm \Gamma}$ there are three singular $q$-points (A, B and C), 
which correspond to $2k_{\rm F}$ or $2(G-2k_{\rm F})$ in the 
noninteracting system indicated by arrows. 
In contrast to half filling, however, the increment at these singular 
points is fairly small.\cite{notepeak} 
We plot $1/N(q)$ at these points versus $\gamma$ for three system sizes 
in Fig.~\ref{fig:cfqfinf}. 
Although system-size dependence is not simple, it is certain that 
$N(q)$ converges even at $\gamma=0$ for each of the three singular 
wave numbers. 
Thus, the CDW correlation in GWF is not intrinsic even at a quarter 
filling, where the band filling is quasi commensurate. 
On the other hand, $P(q)$ in Fig.~\ref{fig:cfqf}(b) is still greatly 
enhanced at the ${\rm \Gamma}$ point by attractive correlation. 
As shown in Fig.~\ref{fig:cfqfinf}, the behavior of $P(0)$ for 
less-than-half filling is essentially the same with that for $n=1$ 
[Fig.~\ref{fig:CFinf}(a)]; although $P(0)$ diverges in the limit of 
$\gamma\rightarrow 0$ and $L\rightarrow\infty$, a long-range order 
of singlet pairing is not realized. 
The value of $U$ at which appreciable system-size dependence appears 
is approximately $U_{\rm co}$ ($\gamma\sim0.1$). 
Consequently, the behavior of $N(q)$ and $P(q)$ of GWF basically 
matches up to the accepted phase diagram of AHM. 
\par

\begin{figure}
\epsfxsize=6cm
\epsfysize=2cm
\caption{
Inverse of $N(q)$ at three singular wave numbers $q=q_{\rm A}$, 
$q_{\rm B}$ and $q_{\rm C}$ and inverse of $P(q)$ at $q=(0,0)$ as 
a function of $\gamma$ for three system sizes at a quarter filling. 
Concerning $P(q)$, the values at half filling are also shown for 
comparison. 
The singular wave numbers of $N(q)$ are incommensurate and subtly 
different size by size; thereby the system-size dependence of $N(q)$ 
is more complicated. 
}
\label{fig:cfqfinf}
\end{figure}
\par

\section{Jastrow-type long-range correlation factor} 
In this Appendix as an alternative prescription to improve the 
normal state we introduce a distance-dependent correlation factor. 
This orthodox Jastrow-type wave function was useful for RHM in 
$n<1$.\cite{YS3} 
We describe the wave function in \S C.1, and discuss the VMC results 
in \S C.2.
\par 

\subsection{Jastrow-type correlation factor} 
It is natural that even in the model with only onsite interaction 
the influence of interaction is not confined to the same site. 
In RHM the importance of repulsive intersite correlation factors 
were obvious, because configurations in which electrons are mutually 
apart are favorable for both kinetic and potential terms. 
Nevertheless, in AHM the situation is not so simple, because electrons 
in proximity seems, if anything, advantageous to the potential term 
but at a glance disadvantageous to the kinetic term. 
The aim of this section is to study how a simple long-range factor 
works in such a system. 
\par

In this work we adopt a two-body Jastrow-type wave function with a 
spin-independent correlation factor:
\begin{equation}
\Psi_{\rm J}={\cal P}_{\rm J}\Phi_{\rm F}
=\prod_{i>j,\sigma\tau}
\left\{1-\left[1-\eta({\bf r}_{ij})\right]n_{i\sigma}n_{j\tau}\right\}
\Phi_{\rm F}, 
\label{eq:Jf} 
\end{equation}
where ${\bf r}_{ij}={\bf r}_i-{\bf r}_j$.
For $\eta({\bf r})$ we employ one of the simplest forms: 
\begin{equation}
\eta({\bf r})=\left\{
\begin{array}{ll}
g & ({\bf r}=0)\\
\left[\sqrt{\sin^2(\frac{\pi}{L}x)+
                  \sin^2(\frac{\pi}{L}y)}\right]^\nu & ({\bf r}\ne 0)
\label{eq:Jfcf} 
\end{array} 
\right., 
\end{equation} 
with ${\bf r}=(x,y)$. 
Such a form was originally introduced to describe a Tomonaga-Luttinger 
liquid for RHM and the $t$-$J$ model in 1D,\cite{HellbergM} and later 
extended to 2D.\cite{GrosV} 
Nevertheless, we overlook the subtle difference in long-range behavior 
between Tomonaga-Luttinger and Fermi liquids,\cite{YO1} because our concern 
here is how a distance-dependent correlation factor works. 
The wave function $\Psi_{\rm J}$ with eq.~(\ref{eq:Jfcf}) has two 
variational parameters $g$ and $\nu$. 
The former one $g$ or $\gamma\ (=1/g)$ controls double occupancy as 
that in ${\cal P}_{\rm G}$, and encourages (discourages) the increase 
of $d$ for $\gamma<1$ ($\gamma>1$). 
On the other hand, $\eta(r)$ substantially works as $r^\nu$ in a short 
distance ($r\ll L$), so that $\eta(r)$ plays a repulsive or an attractive 
role for the intersite part according as $\nu>0$ or $\nu<0$. 
When $\nu$ is zero, the intersite part in eq.~(\ref{eq:Jfcf}) becomes 
constant and $\Psi_J$ is reduced to GWF. 
Incidentally, $\Psi_{\rm J}$ for $U<0$ is not simply connected to 
the counterpart for $U>0$ through eq.(\ref{eq:canon}) even at half 
filling\cite{noteJf} unlike GWF and $\Psi_{\rm Q}$. 
\par

An additional interest in $\Psi_{\rm J}$ is whether a phase separation 
or a CDW arises in AHM. 
For, when this type of wave functions is applied to the $t$-$J$ 
model,\cite{HellbergM,GrosV,YO1} which has an attractive parameter $J$, 
phase separation is brought about as a first-order transition at a finite 
value of attractive interaction, which is quantitatively accurate as 
compared to other reliable results. 
\par

\subsection{Results of VMC calculations}
Now we start with the half-filled band. 
In Fig.~\ref{fig:Jeteu} expectation values of the energy components 
are shown for a variety of parameter sets. 
At first let us look at the case of attractive onsite correlation 
($\gamma<1$).
As for the kinetic part, $|E_t|$ is suppressed for strong attractive 
values of $\nu$ ($\lsim -0.3$) as anticipated and rather unexpectedly 
for $\nu>0$ on account of the tendency toward a fixed configuration 
with the maximum interparticle distance. 
Nevertheless, a weak attractive intersite factor ($-0.25\lsim\nu<0$) 
is somewhat advantageous to the kinetic term. 
This trend stems from the hopping between double occupied sites and 
neighboring empty ones induced by the introduced attractive intersite 
factor.
This is supported by the behavior of $d\ (=E_U/U)$, which is slightly 
reduced for the corresponding small negative values of $\nu$. 
On the other hand, for large negative values of $\nu$, $d$ is enhanced 
as expected. 
Furthermore, note that repulsive intersite correlation also promotes 
double occupation. 
\par

\begin{figure}
\epsfxsize=6cm
\epsfysize=2cm
\caption{
Expectation values of (a) kinetic and (b) potential energies of 
$\Psi_{\rm J}$ at half filling as a function of onsite correlation 
parameter $\gamma$ for some values of intersite correlation parameter 
$\nu$. 
In (b) data for $\nu=-0.1$ and $\nu=-0.25$ almost overlap. 
To survey a wide range of parameters, data of a small system $L=6$ are 
shown; there is no qualitative difference from larger systems. 
$10^6$-$2\times10^6$ samples are averaged for each parameter set. 
Statistical errors are much smaller than the symbol size. 
}
\label{fig:Jeteu}
\end{figure}

In Fig.~\ref{fig:Jetot} total energy $E/t$ is plotted for three values 
of $U/t$. 
Figure \ref{fig:Jetot}(a) shows a typical case of small $|U|/t$, 
where the energy minimum is given by $(\gamma, \nu)=(0.524,-0.05)$. 
This optimal state is realized by small attractive intersite correlation 
with a compensatory enhancement in the attractive onsite part. 
Since the improvement in energy on GWF ($0.548, 0$) is very small 
as seen in the inset, this state is no more than a weakly perturbed GWF.
In this case $E/t$ monotonically increases with $\nu$'s going away 
from the optimal value $-0.05$. 
Figure \ref{fig:Jetot}(b) shows the case for $U/t=-6.8$, in which 
the minimum is given by a similarly small negative value of $\nu$ 
$(0.314,-0.05)$ indicated by `Metal', therefore the realized 
state is still a weakly perturbed GWF. 
Nevertheless, a distinct feature from (a) shows up; $E/t$ becomes 
non-monotonic with respect to $\nu$ and there appears another local 
minimum at (0.851,0.75) indicated by `CDW'. 
These parameter values are quite different from those of `Metal', namely 
weak onsite attraction and fairly strong intersite repulsion. 
Depicted in \ref{fig:Jetot}(c) is the total energy for a slightly 
stronger value $U/t=-7$, where the optimal parameter set is switched 
to `CDW' (0.952,0.83) from `Metal' at (0.299,-0.05). 

\begin{figure}
\epsfxsize=6cm
\epsfysize=2cm
\caption{
Expectation values of total energy at half filling for three values 
of $U/t$, (a) $-4$, (b) $-6.8$ and (c) $-7$. 
Inset in (a) is the magnification of the minimum part. 
Arrows indicate minima, particularly in (b) and (c) large (small) 
arrows means the global (local but not global) minima. 
`Metal' and `CDW' indicate the attributes of the minima. 
Symbols are common to all the panels. 
}
\label{fig:Jetot}
\end{figure}

In Fig.~\ref{fig:Jpara} we show the optimized parameters for 
$0\ge U/t\ge -32$ and for three values of $L$.
For $|U|/t\lsim 6$, only onsite attractive correlation grows with 
$\nu$ keeping almost a constant value $-0.05$. 
At $U/t=U_{\rm CDW}/t\sim -6$, however, it switches to a distinct 
curve, along which both correlation factors become more repulsive 
with increasing $|U|/t$, and sizable system-size dependence appears. 
Thus, some first-order transition is anticipated at this critical value. 
Recall that the intersite repulsive factor enhances double occupation 
[Fig.~\ref{fig:Jeteu}(b)], but obviously hates a gathering configuration 
like phase separation. 
In fact, a transition arising here is from metallic to ($\pi,\pi$)-CDW, 
as will be identified below. 
It is interesting that in the model with only attractive interaction 
the state can be stabilized by the competition among repulsive 
correlations. 

\begin{figure}
\epsfxsize=6cm
\epsfysize=2cm
\caption{
Optimal parameter values of $\Psi_{\rm J}$ at half filling for three 
system sizes. 
Due to discrete parameter values used, the data sometimes show zigzags. 
Some of the corresponding values of $U/t$ is shown. 
The value of $|U_{\rm CDW}|/t$ decreases a little with increasing $L$. 
Inset shows energy improvement by $\Psi_{\rm J}$ on GWF as a function 
of $U/t$ at half filling for three system sizes. 
In the thermodynamic limit the maximum value of $\Delta E/t$ at half 
filling (at $U/t\sim -12$) is roughly extrapolated as $0.26$.
For $|U|>|U_{\rm co}|$, $\Delta E/t$ is approximately proportional 
to $t/|U|$.
}
\label{fig:Jpara}
\end{figure}

Before discussing the correlation function, we add a couple of 
notes on energy. 
First, by studying the optimized energy components (not shown), 
one finds that both $E_t/t$ and $E_U/U$ have such jumps at $U=U_{\rm CDW}$
as $|E_t/t|$ decreases but $|E_U/U|$ increases, in accordance with  
ordinary phase transitions to some ordered states. 
Second, as depicted in the inset of Fig.~\ref{fig:Jpara} the energy gain by 
$\Psi_{\rm J}$ remains very little for $|U|<|U_{\rm CDW}|$ 
[Fig.~\ref{fig:Jetot}(a)], whereas $\Delta E$ abruptly increases 
for $|U|>|U_{\rm CDW}|$. 
This contrasts with a mean-field-type CDW wave function,\cite{Y} 
in which a transition takes place at $U=0$, and thence $\Delta E$ 
is smooth. 
Third, the maximum of $\Delta E$ is situated at $U\sim U_{\rm co}$ of 
GWF, and the system-size dependence of $\Delta E$ is conspicuous for
$|U|>|U_{\rm CDW}|$. 
They are mostly attributed to GWF, as we have experienced for $\Psi_Q$. 
\par

Now, we proceed to the correlation functions.
As shown in Fig.~\ref{fig:JCF}(a), momentum distribution $n(k)$ 
[eq.~(\ref{eq:nksq})] for $|U|<|U_{\rm CDW}|$ represents a typical 
metallic state with discontinuity at $k_{\rm F}$. 
In this wave function an unphysical feature of GWF described in 
Appendix B is remedied by a small but effective intersite correlation 
factor $\mu=-0.05$. 
On the other hand, since the discontinuity at $k_{\rm F}$ almost 
disappears for $|U|>|U_{\rm CDW}|$, the state becomes insulating. 
Next, from $S(q)$ in Fig.~\ref{fig:JCF}(b) 
it is found that spin correlation is suppressed especially for 
$|U|>|U_{\rm CDW}|$, which reflects the sudden increase of onsite 
spin pairs. 
In the limit of $|q|\rightarrow 0$, $S(q)$ seems a linear function of 
$|q|$ even for $|U|>|U_{\rm CDW}|$. 
Therefore, $\Psi_{\rm J}$ does not have a gap in the spin degree of 
freedom. 
Concerning onsite singlet-pair correlation [Fig.~\ref{fig:JCF}(d)],
the sharp peak at $q=0$, which is enhanced with increasing $|U|$ for 
$|U|<|U_{\rm CDW}|$, is suddenly suppressed at $U=U_{\rm CDW}$, and 
$P(q)$ approaches a constant $0.5$ for $|U|>|U_{\rm CDW}|$. 
This indicates that for $|U|>|U_{\rm CDW}|$ pairing correlation is 
restricted to a very short range, especially onsite. 
\par

$N(q)$ [Fig.~\ref{fig:JCF}(c)] increases as a whole with increasing 
$|U|$ in the weak correlation phase ($|U|<|U_{\rm CDW}|$), whereas it 
decreases for $|U|>|U_{\rm CDW}|$ except for $q=G$; $N(G)$ becomes 
overwhelmingly large in the strong-coupling phase, suggesting an 
antiferro CDW order in this phase. 
To confirm the formation of a long-range order, we plot $N(G)$ versus 
$|U|/t$ for a couple of system sizes in the inset of Fig.~\ref{fig:JCF}(c). 
For $|U|>|U_{\rm CDW}|$, $N(g)$ is proportional to the system size, 
and approaches the value of completely polarized CDW state, that is $N$, 
as $|U|$ increases. 
In fact, antiferro CDW patterns are evidently seen in all the snapshots 
taken during Monte Carlo sweeps in the strong correlation phase. 
Thus, we have positive proof of a CDW order. 
Incidentally, one observes in Fig.~\ref{fig:JCF}(c) that $N(q)$ in the 
CDW phase is not a linear function of $q$ around the ${\rm \Gamma}$ 
point but $q^\beta (\beta>1)$.
Thus, a gap opens in the charge sector. 
\par

\begin{figure}
\epsfxsize=6cm
\epsfysize=2cm
\caption{
(a) Momentum distribution and correlation functions of (b) spin, 
(c) charge density and (d) onsite singlet pairing, calculated with 
$\Psi_{\rm J}$ at half filling. 
Path in the momentum space is basically the same with 
Figs.~\ref{fig:GWFcf}(a) and \ref{fig:cflthf}. 
For this system ($L=10$) $U_{\rm CDW}\sim -6.3$. 
The data for $|U|<|U_{\rm CDW}|$ are shown by solid symbols, otherwise 
by open symbols, common in all the panels. 
For each value of $U/t$, $5\times 10^5$ samples are used. 
Inset of (c) shows $N(q)$ at $q=G=(\pi,\pi)$ as a function 
of coupling strength for three system sizes.
Crosses with arrows on the right vertical axis indicate the values for 
the completely polarized CDW: $N$. 
In addition, the order parameter $O_{\rm CDW}$ for $L=10$ is drawn by 
a dashed line.\cite{noteCDWfl}
}
\label{fig:JCF}
\end{figure}

Finally, we look at less-than-half filling, where antiferro CDW 
correlation loses its advantage, because the nesting condition is no 
longer satisfied. 
As for $E_t$ and $E_U$, the behavior is analogous to that at half 
filling (Fig.\ref{fig:Jeteu}); weak attractive intersite correlation 
and proper attractive onsite correlation slightly stabilize the energy. 
A clear difference from half filling is that such a Fermi-liquid state 
is stable even for very large values of $|U|/t$ at least up to $32$ 
and a CDW phase does not appear. 
In Figs.~\ref{fig:Jetotpf}(a) and (b) energy expectation values for 
$U/t=-16$ are depicted for $n=0.72$ and $0.24$, respectively. 
In contrast to Fig.~\ref{fig:Jetot}(c) the minimum of `CDW' never shows 
up in both densities.\cite{noteJqf} 
Thus, we conclude that $\Psi_{\rm J}$ remains a metallic state for all 
the values of $U/t$, but improves GWF only a little. 

\begin{figure}
\epsfxsize=6cm
\epsfysize=2cm
\caption{
Expectation values of total energy in partially filling (a) $n=0.72$ 
and (b) $n=0.24$ for a relatively large value of $|U|/t$. 
Arrows indicate energy minima. 
$5\times 10^5$-$10^6$ samples are used for each parameter set. 
}
\label{fig:Jetotpf}
\end{figure}
\par

As a metallic state the improvement by an orthodox Jastrow-type wave 
function is considerably small for AHM regardless of electron density. 
This results sharply contrasts with those for RHM, for which extensive 
improvement is achieved for $n<1$. 
This is mainly because the distance dependence of electron correlation 
in AHM is not monotonic like the correlation factor in $\Psi_{\rm J}$. 
An extension to a many-parameter algorithm\cite{OVMC} is desired. 

\section{Expansion of $\Psi_Q$ near $U_Q$} 

For a further discussion on the metal-insulator transition in $\Psi_Q$, 
it is useful to consider the expansion in $1-\mu$ again 
[see Appendix C of ref.~\citen{YS3}]. 
In the vicinity of $\mu=1$, one can expand $\Psi_Q$ with respect to 
$1-\mu (\equiv m)$ as, 
\begin{equation}
\Psi_Q=\Psi_Q^{(0)}+
            m\Psi_Q^{(1)}+
            m^2\Psi_Q^{(2)}+\cdots,
\label{eq:defexp}
\end{equation}
$$
\Psi_Q^{(0)}=[\prod_i(1-Q_i)]\ \Psi_{\rm G},\qquad 
\Psi_Q^{(1)}=\bigl\{\sum_jQ_j[\prod_{i(\ne j)}(1-Q_i)]\bigr\}\ \Psi_{\rm G},\ 
$$
$$
\Psi_Q^{(2)}=\bigl\{\sum_{j,k(\ne j)}Q_jQ_k[\prod_{i(\ne j,k)}
                                      (1-Q_i)]\bigr\}\ \Psi_{\rm G}.
$$
Thereby, the energy expectation value is expanded, by abbreviating 
$|\Psi_Q^{(\ell)}\rangle$ to $|\ell\rangle$, as
\begin{equation}
E_Q=
\frac{\langle\Psi_Q|{\cal H}|\Psi_Q\rangle}
     {\langle\Psi_Q|\Psi_Q\rangle}=
      E_Q^{(0)}+mE_Q^{(1)}+m^2E_Q^{(2)}+\cdots, 
\label{eq:deEexp}
\end{equation}
$$
E_Q^{(0)}=\langle0|{\cal H}|0\rangle/N_0,\quad 
E_Q^{(1)}=\left[\langle 0|{\cal H}_t|1\rangle+
          \langle 1|{\cal H}_t|0\rangle\right]/N_0,\ 
$$
\begin{equation}
E_Q^{(2)}=\left[\langle 2|{\cal H}_t|0\rangle+
                \langle 1|{\cal H}_t|1\rangle+
                \langle 0|{\cal H}_t|2\rangle+
                \langle 1|{\cal H}_U|1\rangle+
             N_1\langle 0|{\cal H}|0\rangle\right]/N_0,\ 
\label{eq:udE2}
\end{equation}
with $N_\ell=\langle \ell|\ell\rangle$. 
According to eq.~(\ref{eq:deEexp}), if $E_Q$ has a minimum at $\mu<1$, 
namely if $\Psi_Q$ is metallic, $E_Q^{(1)}$ must be negative and finite. 
Actually, we numerically confirmed that $E_Q^{(1)}$ is negative and its 
absolute value is increasing with $L$ for 1D systems;\cite{YS3} 
this is probably true in the present 2D case. 
Thus, there is no metal-insulator transition within $\Psi_Q$ at finite 
$U$ in the exact sense. 
\par

Leaving aside this fact, we discuss the relevance of the leading 
order $\Psi_Q^{(1)}$ to the transition. 
$\Psi_Q^{(1)}$ consists only of such configurations as 
$A=(\cdots\bigcirc\uparrow\ \downarrow\ \uparrow\bigcirc
\downarrow\bigcirc\cdots)$, in which at least one $\sigma$-spin site 
is adjacent to {\it more than one} $-\sigma$-spin sites ($\bigcirc$ 
indicates either d-site or e-site). 
This configuration is included in a higher-order contribution in 
$\gamma$ expansion than ones belonging to $\Psi_Q^{(2)}$ like 
$B=(\cdots\bigcirc\bigcirc\ \bigcirc\uparrow\bigcirc\downarrow
\bigcirc\cdots)$, 
so that with decreasing $\gamma$ the weight of $\Psi_Q^{(1)}$ becomes 
smaller by order of $\gamma$ than that of $\Psi_Q^{(2)}$. 
Thus, the absolute value of $E_Q^{(1)}$ may become negligibly small 
as compared with $E_Q^{(2)}$, to which lower-order terms in $\gamma$ 
contribute [the last term in eq.(\ref{eq:udE2})]. 
Actually, a conspicuous reduction of $|E_Q^{(1)}|$ with decreasing $\gamma$ 
is seen for a 1D system in Fig.~20 of ref.~\citen{YS3}. 
Thus, the leading term in $E_t/t$ becomes practically 
quadratic in the $\mu\rightarrow 1$ limit for relatively small values 
of $\gamma$ [\eg $\gamma=0.25$ in the inset of Fig.~\ref{fig:udetu}(a)], 
in contrast to the linear behavior for large $\gamma$. 
Recall that the transition in $\Psi_Q$ takes place at $\gamma\sim 0.2$ 
(Fig.~\ref{fig:udparam}). 
Consequently, the problem of our concern virtually rests on $E_Q^{(2)}$. 
Actually, as shown in Fig.\ref{fig:udetot}, total energy $E/t$ for 
large $|U|/t$ does not look linear in $1-\mu$ near $\mu=1$ but becomes 
very flat as a function of $\mu$, and the optimized energy given by 
$\mu<1$ is almost the same with the insulating case $\mu=1$. 
\par

In conclusion, since the contribution of $E_Q^{(1)}$ is irrelevant 
at least for small values of $\gamma$, a substantial metal-insulator 
transition can take place even in $\Psi_Q$. 
Nevertheless, a metal-insulator transition will be treated more clearly, 
if we further improve the wave function by allowing for the following 
point. 
Namely, in the standpoint of $t/U$ expansion all the configurations 
in $\Psi_Q^{(\ell)}$ with odd $\ell$ should be classified in the 
higher-order terms than $\ell$-th. 
For example, the above configuration $A$ ($\ell=1$) should be assigned 
to the same order with the one $B$ ($\ell=2$). 
\par


\begin{thebibliography}{99}
\bibitem{AI} A.J.~Leggett, {\it Modern Trends in the Theory of 
Condensed Matter} (Springer-Verlag, 1980), p.13, 
P.~Nozi\`eres and S.~Schmitt-Rink, \journal{\JLTP}{59}{195}{1984}.
\bibitem{Micnus} An early review is given by R.~Micnus, J.~Ranninger 
and S.~Robaszkiewicz, \journal{\RMP}{62}{113}{1990}.
\bibitem{Scalettar} R.T.~Scalettar \etal, \journal{\PRL}{62}{1407}{1989}.
\bibitem{TR} M.~Randeria \etal, \journal{\PRL}{69}{2001}{1992},
N.~Trivedi and M.~Randeria, \journal{\PRL}{75}{312}{1995}.
\bibitem{Singer} J.M.~Singer \etal \journal{\PRB}{54}{1286}{1996}. 
\bibitem{TM} S.~Schmitt-Rink, C.~Varma and A.E.~Ruckenstein, 
\journal{\PRL}{63}{445}{1989}, R.~Fr\'esard, B.~Glaser and P.~W\"olfle, 
\journal{\JPCM}{4}{8565}{1992}, R.~Micnus \etal, 
\journal{\PRB}{52}{16223}{1995}.
\bibitem{MetznerDMF} M.~Keller, W.~Metzner and U.~Schollw\"ock, 
\journal{\PRB}{60}{3499}{1999}, \journal{\PRL}{86}{4612}{2001} 
and preprint (cond-mat/0109343). 
\bibitem{Capone} M.~Capone, C.~Castellani and M.~Grilli,
preprint (cond-mat/0109194). 

\bibitem{noteexp} This result exhibits a similar tendency to an 
experiment for La systems; in low-temperature normal phase, which 
was realized by destroying the superconductivity in strong magnetic 
field, resistivity is insulating in the underdoped regime but 
metallic in the overdoped one. See G.S.~Boebinger \etal: 
\journal{\PRL}{77}{5417}{1996}.

\bibitem{Guber} M.~Guerrero, G.~Oritz and J.E.~Gubernatis, 
\journal{\PRB}{62}{600}{2000}.
\bibitem{Canon} K.~Dichtel, R.J.~Jelitto and H.~Koppe,
\journal{\ZP}{246}{248}{1971}.
\bibitem{Shiba} H.~Shiba, \journal{\PTP}{48}{2171}{1972}. 
\bibitem{Nagaoka} Y.~Nagaoka, \journal{\PTP}{52}{1716}{1974}.
\bibitem{Harris} A.B.~Harris and R.V.~Lange, \journal{\PR}{157}{295}{1967}.
\bibitem{Emery} V.J.~Emery, \journal{\PRB}{14}{2989}{1976}.
\bibitem{YS1} H.~Yokoyama and H.~Shiba, \journal{\JPSJ}{56}{1490}{1987}.
\bibitem{GWF} M.C.~Gutzwiller, \journal{\PRL}{10}{159}{1963}.
\bibitem{t-J} C.~Gros, R.~Joynt and T.M.~Rice, \journal{\PRB}{36}{381}{1987}; 
H.~Yokoyama and M.~Ogata, \journal{\JPSJ}{65}{3615}{1996}.
\bibitem{KY} Y.~Kuramoto and H.~Yokoyama, \journal{\PRL}{67}{1338}{1991}.
\bibitem{Gutz} M.C.~Gutzwiller, \journal{\PR}{134}{A1726}{1965}. 
\bibitem{Voll} D.~Vollhardt, \journal{\RMP}{56}{99}{1984}.
\bibitem{BRT} W.F.~Brinkman and T.M.~Rice, \journal{\PRB}{2}{4302}{1970}
\bibitem{MV}
W.~Metzner and D.~Vollhardt, \journal{\PRB}{37}{7382}{1988}.
\bibitem{GV}
F.~Gebhard and D.~Vollhardt, \journal{\PRB}{38}{6911}{1988}.
\bibitem{GAA} G.A.~Medina, J.~Simonin and M.D.~N\'u\~nez Regueiro,
\journal{\PRB}{43}{6206}{1991}.

\bibitem{noteSSD}
System-size dependence is very important in GWF in connection with BRT.
Potential energy $E_U$ has large dependence near $\gamma=0$ 
[Fig.~\ref{fig:EGINF}(b)] but no dependence at $\gamma=1$ regardless of 
the value of $n$. 
On the contrary $E_t$ has the largest dependence at $\gamma=1$. 
Thus, the system-size dependence of total energy $E$ is subject to 
$E_t$ for small $|U|$ but to $E_U$ for large $|U|$. 
At half filling, since $E_t$ for $U/t=0$ is scaled as a linear function 
of $1/L^2$, $E/t$ for small values of $|U|/t$ is well extrapolated 
by $1/L^2$. 
For large $U/t$, although a scaling function is a priori not known, 
$E/t$ is smoothly extrapolated by polynomial fit. 
At a quarter filling the dependence of $E_t$ is not monotonic for 
$U/t=0$.
However, the difference between $L=20$ and $L=28$ is so small that 
we regard these sizes as sufficiently large. 
For large $U/t$ the dependence recovers monotonicity, thus a similar 
extrapolation to half filling is possible. 
These changes in system-size dependence from `small' $|U|$ to `large' 
$|U|$ occurs at $|U|/t\sim 11$, and is closely related to the crossover. 

\bibitem{YS3}
H.~Yokoyama and H.~Shiba, \journal{\JPSJ}{59}{3669}{1990}.

\bibitem{noteGWFcor} 
Generally, by introducing some repulsive interaction, $n(k)$ is modified 
from a step function to a decreasing function of $k$. 
Simultaneously, $N(q)$ is suppressed as a whole and $S(q)$ is enhanced 
especially at $q=2k_{\rm F}$ and $2(G-k_{\rm F})$. 
Such changes become more conspicuous as the electron density approaches 
half filling. 
Concerning GWF, however, there appear unusual aspects common in 1D and 2D. 
For example, $n(k)$ is an {\it increasing} function of $k$ both in 
and outside the Fermi surface at half filling [Fig.~\ref{fig:GWFcf}(a)]; 
in less-than-half filling, although $n(k)$ is still an increasing function 
outside the Fermi surface, it is {\it almost constant} for $k<k_{\rm F}$
[Fig.~2(b) in ref.\citen{YS3}]. 
As for $S(q)$ [$N(q)$], there does not exist a peak [suppression] at 
2$k_{\rm F}$ and $2(G-k_{\rm F})$ for $n<1$ [Fig.~2(a) in ref.\citen{YS3}]. 
These aspects of GWF contradict the results of perturbation expansion 
in the weak correlation limit, and are obviously unfavorable for RHM. 
However, all these drawbacks are remedied by introducing intersite 
correlation factors through $\Psi_{\rm J}$ or $\Psi_Q$. 
\par

\bibitem{noteincomme} This is merely an artifact of coarse $q$ division 
due to the small system size. 
For $L\rightarrow\infty$ singularity ought to deviate from $q=G$ 
as soon as $n$ goes away from 1. 

\bibitem{KHF-Fazekas} T.A.~Kaplan, P.~Horsch and P.~Fulde,
\journal{\PRL}{49}{889}{1982}, 
P.~Fazekas, \journal{Physica Scripta T}{29}{125}{1989}.

\bibitem{noteFazekas} We have also studied a similar wave function, 
$
\Psi_{\rm Q'}=\prod_{j\tau}
\left[1+\mu(s^\uparrow_js^\downarrow_{j+\tau}
           +s^\downarrow_js^\uparrow_{j+\tau})\right]\Psi_{\rm G},
$
which is the transformation of one used for 1D RHM.\cite{KHF-Fazekas,YS3} 
Although this function gives the same variational energy with 
eq.~(\ref{eq:udbinding}) for 2D AHM, the variational parameters $\mu$ 
is involved also with the onsite part, \eg $d\rightarrow 0$ for 
$\mu\rightarrow 1$ as studied in ref.\citen{YS3}. 

\bibitem{YS2}
H.~Yokoyama and H.~Shiba, \journal{\JPSJ}{56}{3582}{1987}.

\bibitem{noteCDW} The SDW wave function (for z component), which was 
taken up in ref.~\citen{YS2}, is transformed by eq.~(\ref{eq:canon}) 
to the CDW wave function. 
On the other hand, the SDW wave function for xy component is transformed 
to the singlet superconducting wave function. [see ref.~\citen{Y}]

\bibitem{notetrans} As discussed in Appendix D, we may regard this 
change of behavior as a phase transition, which is second-order because 
through this change both variational parameters do not have a jump 
but continuously vary. 
This is in sharp contrast with a normal-CDW transition treated in 
Appendix C. 

\bibitem{Bulla} R.~Bulla, \journal{\PRL}{83}{136}{1999}.

\bibitem{noteudnq} Since the isotropy in spin space is preserved also 
for the half-filled $\Psi_{\rm Q}$ under the canonical transformation 
eq.~(\ref{eq:canon}), the relation $N(q)=2P(q-G)$ holds similarly to 
GWF.
Hence, here we concentrate on $N(q)$. 

\bibitem{Laloux} L.~Laloux, A.~Georges and W.~Krauth, 
\journal{\PRB}{50}{3092}{1994}.

\bibitem{noteinstab} We check the energy components $E_t/t$ and $d$ as a 
function of $n$; $d$ is a smooth function of $n$ even for a finite size, 
and its curvature becomes gentle with increasing $|U|$ from quadratic 
for $U=0$ to linear for $|U|=\infty$. 
Meanwhile, $E_t$ at $U=0$ already has weak windings at the corresponding 
densities ($n=0.6$ etc.) to the finite-$U$ cases for finite systems, 
but is deservedly smooth for $L=\infty$. 
For finite $U/t$ this aspect is emphasized, so that $E_t/t$ behaves more 
like straight line segments between the winding points, \eg [0.32, 0.6] 
and [0.6, 1.0] for $L=10$. 
For this reason, $E/t$ becomes more or less convex in those sections of 
$n$, especially for intermediate $U$. 
Although we cannot definitely assert that the finite sizes are 
responsible for the convexity of $E/t$, at least the windings are caused 
by finite system sizes. 
This aspect also arises in GWF. 

\bibitem{notePS} The absence of phase separation in AHM is mentioned 
in some literature.\cite{Moreo} 
In view of the effective Hamiltonian eq.~(\ref{eq:ehamil}) such an extreme 
situation as a complete vacuum and close-packed bosons is not likely to 
occur. 
However, the coexistence of two phases with somewhat different densities 
as DMFT anticipates is not excluded by the present VMC result. 

\bibitem{Moreo} A.~Moreo, D.~Scalapino and E.~Dagotto,
\journal{\PRB}{43}{11442}{1991}. 
\bibitem{Randeria} For instance, M.Randeria, in {\it Proceedings of 
the International School of Physics ``Enrico Fermi"}, ed. G.~Iadonisi, 
J.R.~Schrieffer and M.L.~Chiofalo (IOS, Amsterdam 1998).


\bibitem{MetznerTM} D.~Rohe and W.~Metzner, \journal{\PRB}{63}{224509}{2001}.

\bibitem{Vilk} Y.-M.~Vilk, \journal{\JPCS}{59}{1873}{1998}. 

\bibitem{Santos} R.R.~dos Santos, 
\journal{\PRB}{48}{3976}{1993} and \journal{\PRB}{50}{635}{1994}.

\bibitem{Otsuka} For instance, H.~Otsuka: \journal{\JPSJ}{61}{1645}{1992}.

\bibitem{Y} H.~Yokoyama, in preparation. 

\bibitem{dinf} P.G.J.~van Dongen, F.~Gebhard and D.~Vollhardt, 
\journal{\ZPB}{76}{199}{1989}, 
F.~Gebhard, \journal{\PRB}{41}{9452}{1990}.

\bibitem{Thesis}
H.~Yokoyama, Thesis (University of Tokyo, 1988). 

\bibitem{YT} H.~Yokoyama and S.~Tokizaki, 
\journal{Physica B}{230-232}{418}{1997}.

\bibitem{note1D} Nevertheless, in 1D the smooth extrapolation of raw 
data with sufficiently large $L$ leads to the exact values in the
thermodynamic limit.\cite{Thesis} 
On this analogy singular behavior is not likely to appear for 
$L\rightarrow \infty$ also in 2D and 3D.

\bibitem{notepeak} Small but pointed peaks characterize the singularity 
in this system. 
This behavior is qualitatively different from RHM, as argued in detail 
in \S3.3. 

\bibitem{HellbergM}
C.S.~Hellberg and E.J.~Mele, \journal{\PRL}{67}{2080}{1991}. 
\bibitem{GrosV}
C.~Gros and R.~Valent\'\i, \journal{\MPLB}{7}{119}{1992}.
\bibitem{YO1} H.~Yokoyama and M.~Ogata, \journal{\PRL}{67}{3610}{1991},
and \journal{\PRB}{53}{5758}{1996}.

\bibitem{noteJf} The wave function eq.~(\ref{eq:Jf}) is rewritten as 
$
\Psi_{\rm J}=
\exp\left(\sum_{j\ell\sigma\tau}\alpha_\ell 
n_{j\sigma}n_{j+\ell\tau}\right)\Phi_{\rm F}
$
with $\eta(r_\ell)=\exp(\alpha_\ell)$. 
By applying the canonical transformation eq.~(\ref{eq:canon}), 
$\Psi_{\rm J}$ is transformed to 
$
\tilde\Psi_{\rm J}=
\exp\left[\sum_{j\ell\sigma}\alpha_\ell 
\left(\tilde n_{j\sigma}\tilde n_{j+\ell\sigma} 
     -\tilde n_{j\sigma}\tilde n_{j+\ell-\sigma}\right)\right]
\tilde\Phi_{\rm F}.
$
Accordingly, $\tilde\Psi_{\rm J}$ is not simply connected to the 
original $\Psi_{\rm J}$ even at half filling, because $\tilde\Psi_{\rm J}$ 
has a spin-{\it dependent} Jastrow factor. 

\bibitem{noteCDWfl}
For small systems the order parameter $O_{\rm CDW}$ does not give stable 
values near 1, until $|U|$ becomes somewhat larger than $|U_{\rm CDW}|$. 
For $L=6$ there is no stable region for $|U|/t\le 32$. 
This is caused by the inversion of CDW phase of the whole system 
during the Monte Carlo sweeps. 
As $L$ increases, however, this instability is rapidly suppressed. 
Of course, this instability does not affect two-point correlation 
functions at all.

\bibitem{noteJqf} 
We have also studied a quarter filling ($n=0.5$), where the filling is 
quasi commensurate, although the nesting condition for antiferro CDW 
is not satisfied. 
Thus, a CDW with different wave numbers may appear. 
Actually, a similar transition to half filling takes place at 
$U/t=\sim -12.4$ for $L=8$. 
In the strong coupling side of this transition a CDW state is realized, 
in which the symmetry between $x$ and $y$ axes is broken and $N(q)$ is 
anomalously enhanced at $q=(\pi,0)$ and $(\pi/2,\pi)$. 
However, a long-range CDW order is not observed for larger systems 
with $L=12$ and $20$. 
We suspect that the anisotropic (periodic-antiperiodic) boundary 
condition we employ has a bad influence on that small system size. 
In conclusion a CDW order is absent also for a quarter filling. 

\bibitem{OVMC} C.J.~Umrigar, K.G.~Wilson and J.W.~Wilkins, 
\journal{\PRL}{60}{1719}{1988}, 
K.~Kobayashi and K.~Iguchi: \journal{\PRB}{47}{1775}{1993}.

\end{thebibliography}
\end{document}